\documentclass[aps,prd,showpacs,twocolumn,superscriptaddress,nofootinbib,floatfix,preprintnumbers]{revtex4-1} 

\makeatletter
\def\p@subsection{}
\makeatother

\usepackage{color,graphicx,amsmath,amssymb,lmodern,lipsum,bm}

\usepackage[usenames,dvipsnames]{xcolor}
\definecolor{xlinkcolor}{rgb}{0.7752941176470588, 0.22078431372549023, 0.2262745098039215}

\usepackage[colorlinks=true,citecolor=xlinkcolor,linkcolor=xlinkcolor,urlcolor=xlinkcolor, backref=false,pdfborder={0 0 0}]{hyperref}

\usepackage[normalem]{ulem}
\usepackage[utf8]{inputenc} 
\usepackage{mathtools}

\usepackage{float}
\usepackage{aas_macros}

\usepackage{ulem}
\usepackage{diagbox}

\usepackage[dvipsnames]{xcolor}
\usepackage{mathrsfs}

\definecolor{xlinkcolor}{rgb}{0.7752941176470588, 0.22078431372549023, 0.2262745098039215}

\newcommand{\be}{\begin{equation}}
\newcommand{\ee}{\end{equation}}
\newcommand{\beqa}{\begin{eqnarray}}
\newcommand{\eeqa}{\end{eqnarray}}

\newcommand\p{{\bm p}}
\renewcommand\k{{\bm k}}

\newcommand\x{\bm x}

\newcommand\GG{\Gamma_3}
\newcommand\G{\mathcal{G}_2}

\newcommand{\HH}{{\cal H}}

\newcommand{\bseq}{\begin{subequations}}
\newcommand{\eseq}{\end{subequations}}

\renewcommand{\ln}{\mathop{\rm ln}\nolimits}

\newcommand{\knl}{k_{\rm NL}}

\def\gsim{\raise0.3ex\hbox{$\;>$\kern-0.75em\raise-1.1ex\hbox{$\sim\;$}}}
\def\lsim{\raise0.3ex\hbox{$\;<$\kern-0.75em\raise-1.1ex\hbox{$\sim\;$}}}

\def\beqn#1{\begin{equation}\label{#1}}
\def\eeqn{\end{equation}}

\def\beqa#1{\begin{eqnarray}\label{#1}}
\def\eeqa{\end{eqnarray}}

\def\kmax{{k_\text{max}}}
\def\hMpc{h{\text{Mpc}}^{-1}}
\def\Mpch{h^{-1}{\text{Mpc}}}

\def\Z2{$\mathcal{Z_2}$}

\newcommand {\ignore}[1]{}

\begin{document}

\preprint{MIT-CTP/5851}

\title{Simulation-Based Priors without Simulations: \\
an Analytic Perspective on 
EFT Parameters of Galaxies
}

\author{Mikhail M. Ivanov }
\email{ivanov99@mit.edu}
\affiliation{Center for Theoretical Physics, Massachusetts Institute of Technology, 
Cambridge, MA 02139, USA} 
 \affiliation{The NSF AI Institute for Artificial Intelligence and Fundamental Interactions, Cambridge, MA 02139, USA}

\begin{abstract} 
Effective field theory (EFT)-based full-shape analysis with 
simulation-based priors (SBPs)
is a novel approach to
galaxy clustering data analysis, 
which significantly boosts the constraining power by efficiently incorporating field-level simulation 
information from small scales. 
So far, SBPs have been mostly extracted from a large set of mock catalogs generated with the Halo Occupation Distribution (HOD) approach. 
We show 
that given 
a halo mass function model
and assuming that the EFT parameters of halos depend only on the peak height, 
HOD-based priors can be computed analytically for the standard 7-parameter HOD model. 
We derive the relevant expressions 
for deterministic EFT parameters 
from the halo model. 
The halo model,
however, fails to 
accurately 
describe
stochastic EFT parameters, 
for which we use a physically motivated phenomenological prescription. 
We compare our 
analytic 
priors with simulations and find an excellent agreement. Our  approach provides 
analytic
insights into the physics behind 
the EFT parameters of galaxies, and allows one to reduce the computation time 
of SBPs 
to virtually nothing. 
As an application, we produce a set of 100,000 EFT parameters by sampling both HOD and cosmological models,
and explicitly demonstrate that 
any cosmology dependence can be completely absorbed by 
relatively minor shifts of the HOD parameters. Our approach can be  generalized to other variants of the HOD and cosmological models beyond $\Lambda$CDM. 
\end{abstract}

\maketitle

\section{Introduction}

Effective field theory (EFT) for Large-Scale Structure~\cite{Baumann:2010tm,Carrasco:2012cv,Ivanov:2022mrd}
is a systematic analytic approach 
to structure formation that provides 
unmatched accuracy and flexibility
for quasi-linear clustering observables. 
It has been successfully applied 
to various kinds of cosmological
data in the perturbative regime, e.g. 
galaxy clustering in redshift-space (\cite{Ivanov:2019pdj,DAmico:2019fhj,Chen:2021wdi,Philcox:2021kcw,Chen:2024vuf}, see also~\cite{DESI:2024jis} for the recent 
application of this technique by the Dark Energy Survey Instrument
collaboration),  
projected
galaxy clustering~\cite{Chen:2024vvk},
and the Lyman-$\alpha$ forest~\cite{Ivanov:2023yla,Ivanov:2024jtl,deBelsunce:2024rvv}. 

One important feature of EFT 
is the presence of free EFT parameters
(a.k.a. ``nuisance parameters''), 
which are not calculable 
within the EFT and 
need to be fitted from data
or numerical simulations. 
The standard approach is to determine 
the EFT parameters from the data itself.
While this approach produces conservative 
results (provided that the priors on EFT parameters
are chosen to be appropriately wide),
it leads to a significant degradation
of constraining power, see~\cite{Wadekar:2020hax,Cabass:2022epm,Cabass:2022epm}
for detailed discussions.\footnote{Note that effect is different from the so-called prior projection
effects, originally studied in \cite{Ivanov:2019pdj,Chudaykin:2020ghx,Philcox:2021kcw}, see also~\cite{Chudaykin:2024wlw,Paradiso:2024yqh} for 
recent discussions.
The simulation-based priors
eliminate these 
projection effects~\cite{Ivanov:2024xgb}.
} 

In contrast, 
matching EFT parameters from
the numerical simulations
leads to 
a dramatic improvement 
of constraining power~\cite{Cabass:2024wob,Ivanov:2024hgq,Ivanov:2024xgb}. 
This approach is known as the 
EFT-based full-shape analysis
with the simulation-based priors (SBPs). 
(See also~\cite{Sullivan:2021sof,Zennaro:2021pbe} for similar ideas 
in the context of other clustering models, 
and more recent works~\cite{Zhang:2024thl,Akitsu:2024lyt}.)
The significant information gain can be achieved if the SBPs are calibrated 
at the field level~\cite{Schmittfull:2016jsw,Schmidt:2018bkr,Schmidt:2020ovm,Schmidt:2020tao,Schmidt:2020viy,Nguyen:2024yth, Abidi:2018eyd,Schmittfull:2018yuk,Schmittfull:2020trd}, which allows one
to break the degeneracies present
in the correlation functions~\cite{Ivanov:2024xgb}.
In addition, 
the EFT-based full-shape analysis
with SBPs allows one to 
efficiently incorporate 
small scale information beyond
the commonly used two-point 
clustering observables. 

Until now, SBPs were calibrated
using a combination of N-body simulations and the 
Halo Occupation Distribution (HOD) 
framework~\cite{Berlind:2002rn,Zheng:2004id} to map galaxies onto the
N-body simulation output (see~\cite{Wechsler:2018pic} for a review). 
Ref.~\cite{Ivanov:2024dgv} 
extended this to 
hydrodynamical simulations.
Given a dependence  
of SBPs 
on computationally expensive simulations, 
it is desirable to develop 
an analytic 
approach to SBPs,
which could reduce their 
computational cost. 
In addition, such an approach
should provide insights into 
the physics behind the EFT parameters,
increasing
the interpretability 
of the EFT-SBP results. 
This is especially 
relevant in light of the 
nominal power of this approach
in constraining cosmological parameters~\cite{Ivanov:2024hgq,Ivanov:2024xgb}. 
We address these questions
in our note. 

Our main aim is to use the halo model formalism~\cite{Seljak:2000gq,Seljak:2000jg,Cooray:2002dia} 
to predict the relations between the EFT 
parameters of dark matter halos and 
galaxies. These relations are well
known for the galaxy bias parameters~\cite{Benson:1999mva}.
In this work we critically revisit 
the derivation of these expressions, 
and generalize these calculations 
to the case of redshift-space 
and stochasticity EFT parameters. 
Let us briefly introduce 
the key ingredients of our analytic approach
using the known example of 
the galaxy bias parameters. 

Galaxy and halo bias are 
functions encapsulating the 
dependence of the galaxy and halo overdensity
$\delta_{h/g}$
on the underlying 
dark matter field $\delta_m$ (see~\cite{Desjacques:2016bnm} for a review). 
The bias expansion is formulated 
in terms of operators 
$\mathcal{O}_a(\x)$ built out of the 
dark matter observables such as the tidal field. 
Schematically, the galaxy and halo bias 
expressions can be written as
\be 
\begin{split}
& \delta_h(\x) = \sum_{a=1} b_{\mathcal{O}_a}^h \mathcal{O}_a(\x)\,, \quad  \delta_g(\x) = \sum_{a=1} b_{\mathcal{O}_a}^g  \mathcal{O}_a(\x)\,,
\end{split}
\ee 
where we have omitted the explicit 
time-dependence for brevity. 
The halo model predicts that 
the bias parameters of galaxies
$b_{\mathcal{O}_a}^g$
are given by the 
bias parameters of halos
weighted with the halo mass function $\bar n(M)$, 
and the mean halo occupation 
distribution $\langle N_g\rangle(M)$,
\be 
\label{eq:analyticHOD}
b_{\mathcal{O}_a}^g=\underbrace{\int d\ln M \bar n(M)}_{\text{halo mass function}}~\underbrace{\langle N_g\rangle(M)}_{\text{galaxy-DM connection}}~\underbrace{b_{\mathcal{O}_a}^h(M)}_{\text{Halo bias}}~\,,
\ee 
where the integral above is over the 
halo mass $M$. If we were to compute the bias parameters analytically, we would need three key ingredients. 

The first ingredient is the halo mass 
function.
Thanks to decades of intense theoretical efforts, many analytic results are
readily available
in the literature~\cite{Press:1973iz,Sheth:1999mn,Tinker:2008ff}. 
The second ingredient 
is the halo occupation 
distribution 
$\langle N_g\rangle(M)$
which captures the
galaxy-dark matter
connection. The basic theoretical 
HOD models such as~\cite{Zheng:2004id}
provide analytic formulas
for $\langle N_g\rangle(M)$,
which can be used to do the 
integral~\eqref{eq:analyticHOD}
numerically. Finally, 
the last ingredient 
in the above formula
is the bias parameter 
of halos as a function
of the halo mass, $b_{\mathcal{O}_a}^h(M)$, or a related
quantity of the halo model known as the 
``peak height''~\cite{Seljak:2000gq}. 
Since there are no available 
analytic results for all of the relevant 
EFT parameters for dark matter halos, 
in practice one has to extract
$b_{\mathcal{O}_a}^h(M)$ from  N-body simulations.
However,
these dependencies have to be calibrated
from the simulations only once. 

All in all, with the three ingredients described above, the halo mass function model, 
the analytic expression
for the HOD function,
and the halo bias as a function of the halo 
mass, one can compute 
the HOD-based galaxy bias parameters
analytically.
Our goal here is to 
carry out this computation explicitly,
and
extend this argument
to other EFT parameters.
Finally, we will 
compare the results against numerical simulations. 

The analytic approach described above introduces some uncertainties. First, it relies on an
analytic halo mass
function model, 
which is always an approximation~\cite{Tinker:2008ff,Li:2024wco}.
Second, it uses the analytic form
of the HOD that ignores important 
physical effects such as assembly bias~\cite{Hearin:2015jnf}. 
Third, it assumes that the bias parameters
of halo depends only on the mass, which is known to be violated
in N-body simulations~\cite{Lazeyras:2021dar,Lucie-Smith:2023hil}. 
Nevertheless, despite these drawbacks, our approach 
provides an economic way 
to generate the EFT priors analytically, 
and explore the dependence of the EFT parameters
on the galaxy-halo connection parameters
and the cosmological parameters.
Note that priors do not need to be very
accurate, i.e. the $\sim 10\%$ uncertainties
are acceptable. Given that the currently used
conservative priors allow for $\sim 100\%$
uncertainties on the EFT parameters
and their correlations~\cite{Chudaykin:2020aoj,Chen:2020zjt,Chen:2021wdi}, 
reducing them to $\sim 10\%$
is already a dramatic improvement. 

The rest of our note is structured as follows. 
Section~\ref{sec:theory}
introduces the theoretical background
on the galaxy parameters in the halo model
and presents new expressions for the 
redshift-space and stochastic EFT parameters. 
We compare our analytic results againts
simulations in Section~\ref{sec:sim}.
The dependence of HOD-based priors
on cosmological and HOD parameters 
is studied in Section~\ref{sec:cosmo}.
Section~\ref{sec:disc}
draws conclusions.

\section{Theory}
\label{sec:theory}

Let us describe now three key ingredients of our analytic approach. 
We start with the halo mass function. 

\subsection{Halo mass function}

The halo mass function (HMF) computes the number density 
of halos of a given mass. 
Historically, it was first 
analytically derived by Press and Schechter~\cite{Press:1973iz}. 
Within the Press-Schechter (PS) approach, a halo is formed
at a given point in space  
if the variance of the Lagrangian matter density fluctuation in a spherical cell around this 
point at the initial time slice was above a spherical collapse threshold. 
Using the Lagrangian density formally 
evolved to redshift zero,
its variance inside a  
spherical cell around the point
of interest is given by\footnote{We use 
the notation $\int_\k\equiv \int d^3k$. } 
\be 
\sigma^2(R,z=0)=\int_\k P_{11}(k)W^2_{\rm th}(kR)\,,
\ee 
where $R$ is comoving, i.e. redshift-independent Lagrangian scale, $W^2_{\rm th}$ is the Fourier image of the position space top-hat window of radius $R$, and $P_{11}(k)$
is the linear matter power spectrum at redshift zero. 
The mass variance at a finite redshift $z$ is obtained by multiplying the above
expression by the square of the linear growth factor $D_+(z)$:
\be 
\sigma^2(R,z)=D_{+}^2(z)\sigma^2(R,z=0)\,.
\ee 
The PS formalism assumes that the spherically averaged 
density field in a matter-dominated universe follows a non-linear spherical collapse evolution.~\footnote{This is true for the spherically 
averaged overdensity 
in the statistical sense since the path integral of the spherically averaged matter density 
probability distribution function 
is dominated by a spherical collapse saddle point~\cite{Ivanov:2018lcg}.}
In the spherical collapse model it is customary to use 
the linear theory density field as a proxy for time. 
The collapse of spherical shells (shell crossing) happens
when the linear theory prediction formally reaches the threshold 
value $\delta_c=1.686$,
known as ``critical density.''
Given that the time-evolution is self-similar, shell crossing
at redshift $z$ happens in regions whose linear mass r.m.s. 
$\sigma(R,z)$ crosses the same threshold value $\delta_c$. 
The PS model assumes that shell-crossing is followed by the creation 
of a halo. In this case, the HMF is given by 
\be 
\begin{split}
& \bar n(M)\equiv \frac{d\bar n(M,z)}{d\ln M} 
=\frac{\bar\rho_m(t_0)}{M}\left|\frac{d\ln \sigma(M,z)}{d\ln M}\right|\nu f(\nu)\,,\\
& f(\nu) = \sqrt{\frac{2}{\pi}}\exp\{-\nu^2/2\}\,,
\end{split}
\ee 
where the key parameter that controls the HMF is 
\be 
\nu = \frac{\delta_c}{\sigma(M[R],z)}\,,
\ee 
which we call the ``peak height.''
(Note that~\cite{Seljak:2006bg} uses the same term for  
$\nu^2$.)
The principal dependence of the HMF only on the combination $\nu$
implies \textit{universality}: 
a rescaling of the HMF by redshift can be 
fully 
compensated by a shift of its mass. Consequently,
this implies a self-similar shape of the HMF.

Going beyond the matter dominated universe 
within the PS framework, the cosmology dependence is controlled by
the same combination $\nu$.
Just as before, $\sigma^2$ captures the initial conditions
and the evolution prior to the onset of structure formation, 
while $\delta_c$ is promoted to depend on the time-dependent 
matter abundance 
\be
\Omega_m(z)= \frac{\Omega_{m,0}(1+z)^3}{\Omega_{m,0}(1+z)^3+1-\Omega_{m,0}}\,,
\ee
where $\Omega_{m,0}$ is the matter abundance at $z=0$, as~\cite{Kitayama:1996ne} 
\be 
\delta_c=\frac{3}{20}(12\pi)^{2/3}(1+0.0123\log\Omega_m(z))\,.
\ee

The PS HMF has been generalized in multiple ways. The first popular extension is the  Sheth-Torman mass function (ST HMF)~\cite{Sheth:1999mn}, which accounts for 
the ellipsoidal collapse, 
\be 
f_{\rm ST}(\nu)=A\left[1+\frac{1}{(a\nu)^p}\right]\sqrt{\frac{a}{2\pi}}\exp\{-\nu^2/2\}\,,
\ee 
with $p=0.3$ and $a=0.707$. $A$ is determined through the normalization condition 
that all matter in the Universe is contained in halos,
\be
\label{eq:normST}
\int d\nu ~ f_{\rm ST}(\nu)=1\,.
\ee
Another popular generalization is the Tinker mass function (THMF)~\cite{Tinker:2008ff}, which is a fit to simulations,
\be 
f_{\rm T}(\nu)=\alpha \left[1+(\beta\nu)^{-2\phi}\right]\nu^{2\eta}e^{-\gamma\nu^2/2}\,,
\ee 
with 
\be 
\begin{split}
&\beta = 0.589(1+z)^{0.20}\\
&\phi = -0.729(1+z)^{-0.08}\\
&\eta = -0.243(1+z)^{0.27}\\
&\gamma = 0.864(1+z)^{-0.01}\,,
\end{split}
\ee 
and $\alpha$ is again determined from eq.~\eqref{eq:normST}.
In all these cases the universal shape of the mass function is preserved. 

The universality has been extensively tested with N-body simulations. 
While the corrections have been detected, in general they are very small,
especially for cosmology variations allowed by 
current experiment~\cite{Li:2024wco}. 
In what follows we will assume the universality of the HMF 
and use the THMF in all practical calculations.

\subsection{EFT parameters of halos}

The second ingredient 
of our approach is the set of EFT parameters
of dark matter halos. Let us discuss the various
kinds of EFT parameters one-by-one. 

\subsubsection{Bias parameters}

The EFT Eulerian bias model 
relevant for the 
description of the 
halo power spectrum 
at the one-loop order reads~\cite{Assassi:2014fva,Desjacques:2016bnm}, 
\be 
\label{eq:naive_eft}
\delta^{\rm EFT}_g(\k) = b^h_1\delta + \frac{b^h_2}{2}\delta^2 +b^h_{\mathcal{G}_2}\mathcal{G}_2
+b^h_{\Gamma_3}\Gamma_3 - b'^h_{\nabla^2\delta}\nabla^2\delta
+\epsilon\,,
\ee 
where $\delta$ is the non-linear 
matter density field, 
$\mathcal{G}_2$ is the tidal operator, 
\be 
\label{eq:G2}
\begin{split}
\mathcal{G}_2(\k) & = \int_{\bm p} 
F_{\mathcal{G}_2}(\p,\k-\p)
\delta({\bm p})\delta(\k-{\bm p})\,,\\
 F_{\mathcal{G}_2}(\k_1,\k_2) & =\frac{({\bm k_1}\cdot \k_2)^2}{k_1^2 k_2^2}-1\,,
\end{split}
\ee 
$\int_{\k}\equiv \int \frac{d^3\k}{(2\pi)^3}$
and $\GG$ is the Galileon cubic tidal operator, 
\be 
\label{eq:gamma3def}
\begin{split}
&\GG =  \int_{\k_1}\int_{\k_2}\int_{\k_3}\left(\prod_{i=1}^3\delta(\k_i)\right)(2\pi)^3\delta_D^{(3)}(\k-\k_{123})F_{\GG}\,,\\
& F_{\GG}=\frac{4}{7}\left(1-\frac{(\k_1\cdot\k_2)^2}{k_1^2k_2^2}\right)
\left(\frac{((\k_1+\k_2)\cdot \k_3)^2}{(\k_1+\k_2)^2k_3^2}-1\right)\,.
\end{split}
\ee 
The field $\epsilon$ in eq.~\eqref{eq:naive_eft}
is a stochastic density component.

In perturbation theory~\cite{Bernardeau:2001qr} one expands the non-linear matter field
$\delta$ in expressions above through the linear density  field,
schematically
\be 
\delta = \delta_1+\int F_2 \delta_1^2+\delta_{\rm ctr.}+...\,.
\ee  
$\delta_{\rm ctr.}=-c_s^2k^2\delta_1$ above is the counterterm contribution
needed to account for the backreaction from small scales. 
At leading order it produces a contribution indistinguishable 
from the higher-derivative bias, 
so in what follows it will be convenient to re-define
\be
b^h_{\nabla^2 \delta} = b'^h_{\nabla^2\delta} -b_1 c_s\,,
\ee
as this is the combination that appears in all the calculations
at the one-loop power spectrum order
that we use here.

The bias expression~\eqref{eq:naive_eft} describes the correlation
of the halo overdensity with the underlying matter field. 
Let us discuss briefly now how this expression appears in the halo model~\cite{Desjacques:2016bnm}.
The formation of halos is modulated by long-wavelength modes,
which act as a background on the proto-halos. This leads to correlations
of the positions of halos and the cosmological fluctuations on large scales, 
captured by the local in the density field bias relation
\be 
\label{eq:local_gar_bias}
\delta^L_h(\x_L)= \frac{n(\x_L)}{\bar n}-1 = b^L_1\delta_m(\x_L)+\frac{b^L_2(\x_L)}{2}\delta^2_m+...\,,
\ee 
where $\x_L$ is the Lagrangian coordinate of the initial time slice.
At the linear level this leads to the Eulerian bias relation 
\be 
\delta_h(\x,z) = b^h_1\delta_m(\x,z)
\ee 
where $b^h_1=b_1^L+1$ is the linear bias parameter. In the simplest models
of the halo formation, such as the peak-background split, 
$b_1$ depends only $\nu$ by virtue of the HMF universality. 
Analytic results have been derived for the linear bias
in the context of the PS and ST HMFs~\cite{White:2014gfa}:
\be 
\begin{split}
& b^{\text{PS}}_1(\nu)=1+\frac{\nu^2-1}{\delta_c}\,,\\
& b^{\text{ST}}_1(\nu)=1+\frac{1}{\delta_c}\left(a\nu^2-1+\frac{2p}{1+(a\nu^2)^p}\right)\,.
\end{split}
\ee 
Note that we have added the unity to the above expressions
to convert the Lagrangian bias predicted by the peak background split
model into the observationally relevant Eulerian bias. 
For the THMF, the fit to simulation data reads~\cite{Tinker:2010my}:
\be 
b_1(\nu)=1-A\frac{\nu^a}{\nu^a+\delta_c^a}+B\nu^b + C\nu^c\,,
\ee 
where $y=\log(200)$ and 
\be
\begin{split}
&A =1+0.24 y\exp(-(4/y)^2)\,,\\ 
&a=0.44y-0.88\,,\\
&B=0.183\,,\\
&b=1.5\,,\\
&C=0.019+0.107y+0.19\exp(-(4/y)^2)\,,\\
&c=2.4\,.
\end{split}
\ee
We note that this fit was calibrated to reproduce the linear
bias in the range $\ln\nu\in [-0.3,0.5]$ within $\lesssim 20\%$. 
Analogous expressions have been obtained for higher order
local in density field bias parameters, e.g. 
\be 
\label{eq:b2ST}
\begin{split}
& b^{\text{ST}}_2(\nu)\big|_{L}=\frac{1}{\delta^2_c}\left(a^2\nu^4-3a\nu^2+\frac{2p(2a\nu^2+2p-1)}{1+(a\nu^2)^p}\right)\\
& b_2=b_2\big|_L+\frac{8}{21}(b_1-1)+\frac{4}{3}b_{\G}=b_2\big|_L
+\frac{4}{3}b_{\G}
\big|_L
\,,
\end{split}
\ee
where $b_{\G}
\big|_L$ is the Lagrangian
tidal bias. 
The PS prediction can be recovered by setting $a=1$, $p=0$.

Within the local-in-the-density-field Lagrangian bias model 
the Lagrangian bias is assumed to be local cf.~\eqref{eq:local_gar_bias}, as follows
from the PBS arguments. All other terms in 
the EFT expression~\eqref{eq:naive_eft} are generated
by the mapping from Lagrangian to Eulerian spaces. In particular, 
the tidal bias satisfies 
\be
\label{eq:bG2LocL}
b_{\G} = -\frac{2}{7}(b_1-1) \,.
\ee
This picture 
is known to be oversimplified, as 
the Lagrangian halo bias in simulations also depends
on the tidal fields, as well as the higher derivative terms~\cite{Abidi:2018eyd,Eggemeier:2021cam}.

In the basic formulation of the  halo model the bias parameters
of halos depend only on the halo mass though the parameter $\nu$.
This parameter also captures the cosmology dependence. 
Within this framework, 
it is natural to expect then that other bias parameters, 
such as $b_{\G}$, $b_{\Gamma_3}$, $b_{\nabla^2\delta}$ etc. primarily depend
on the masses and cosmology through peak height $\nu$ 
as well. 
Since the analytic results for these bias parameters as functions
of $\nu$ are not available, we will use the dependencies
extracted from simulations. This applies to $b_1(\nu)$
and $b_2(\nu)$ as well, 
as in the analytic formulas presented above work only within $\sim 10\%$
precision~\cite{Lazeyras:2015lgp}.

\begin{figure*}
\centering
\includegraphics[width=0.49\textwidth]{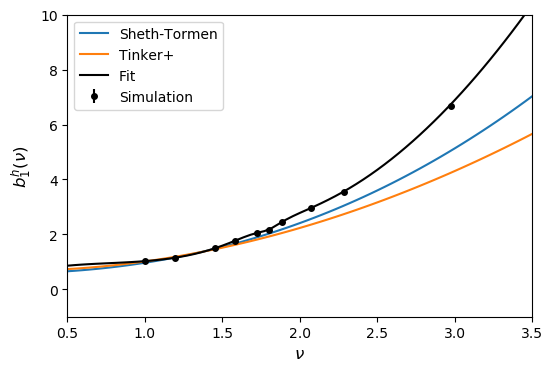}
\includegraphics[width=0.49\textwidth]{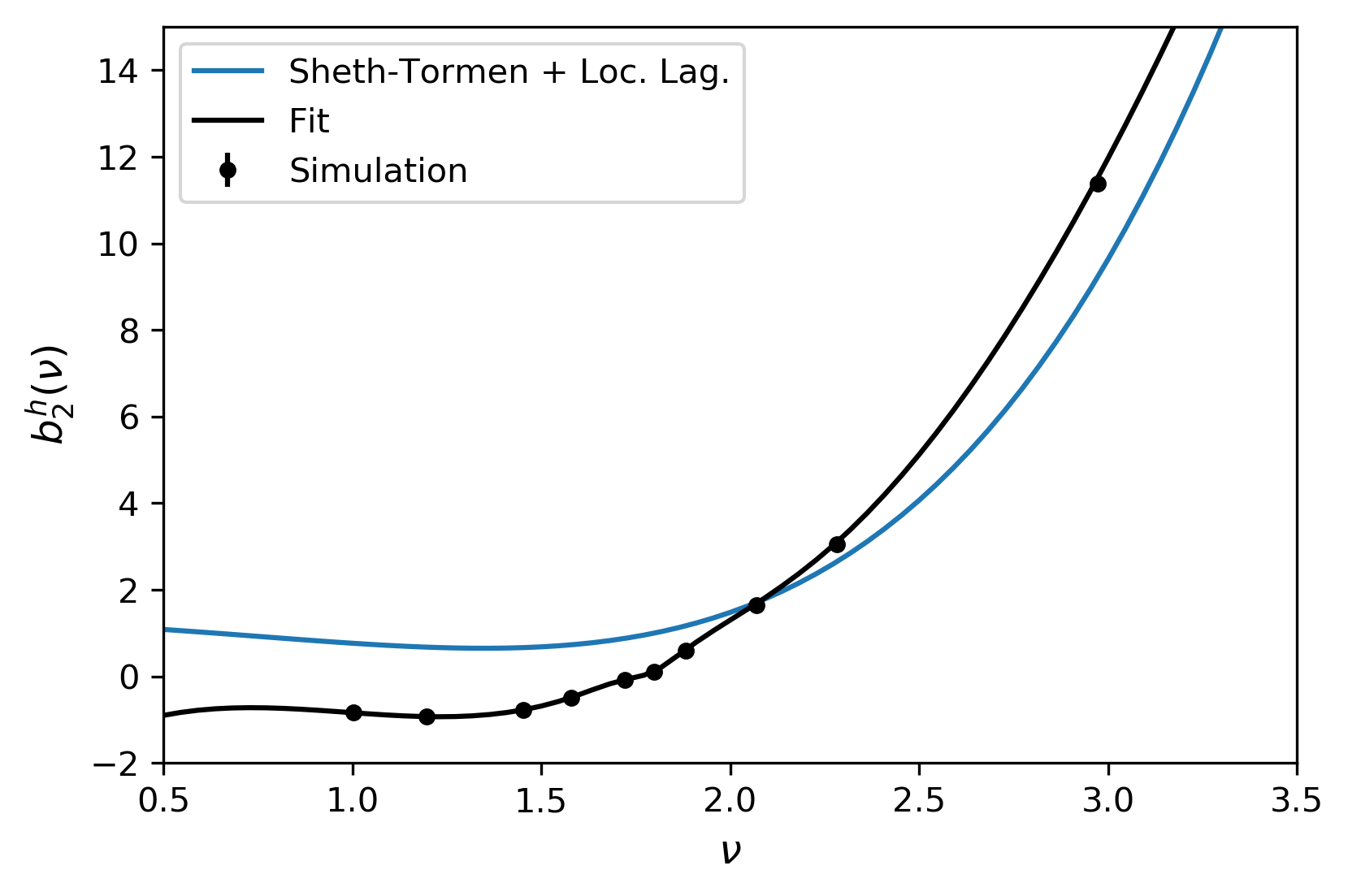}
\includegraphics[width=0.49\textwidth]{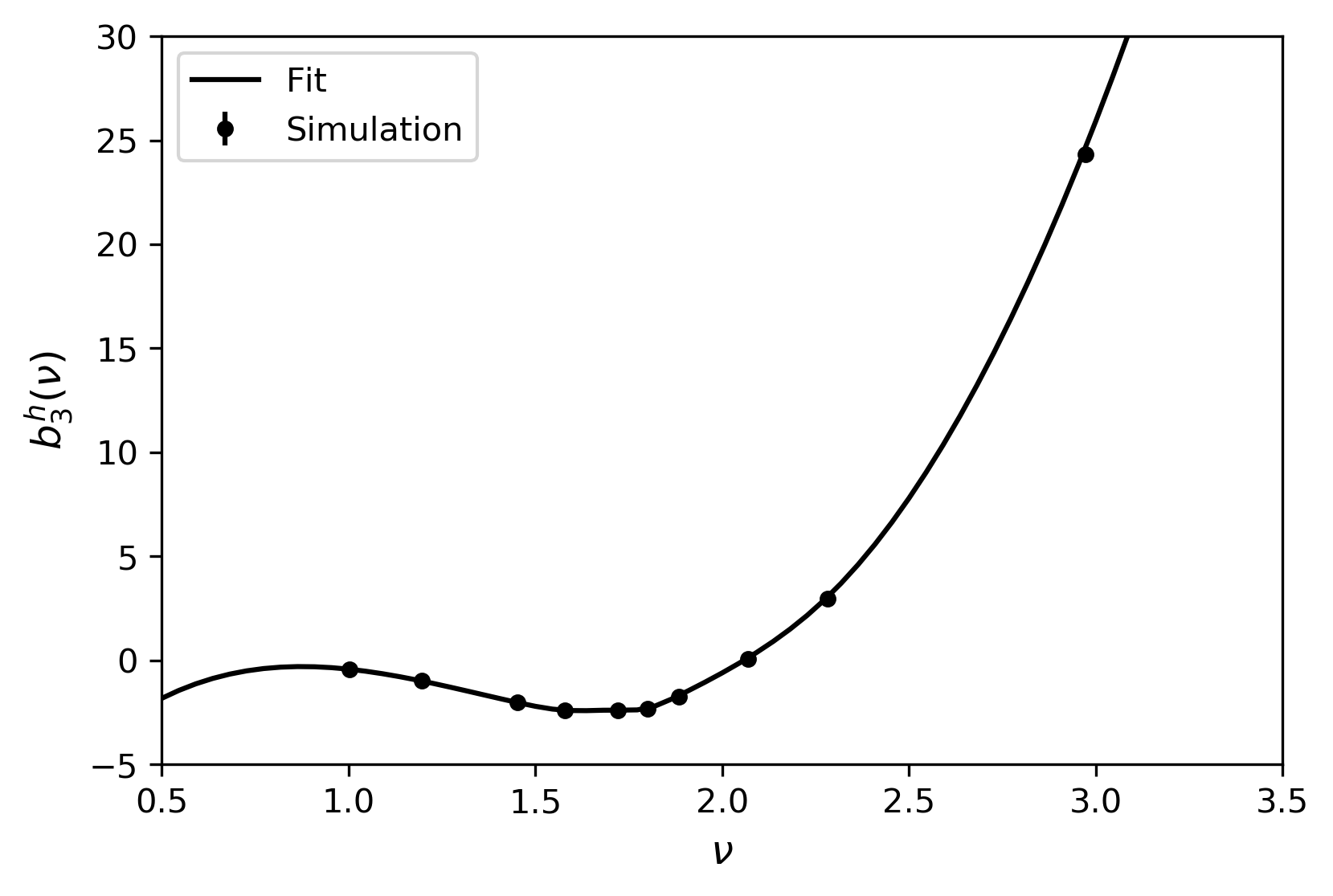}
\includegraphics[width=0.49\textwidth]{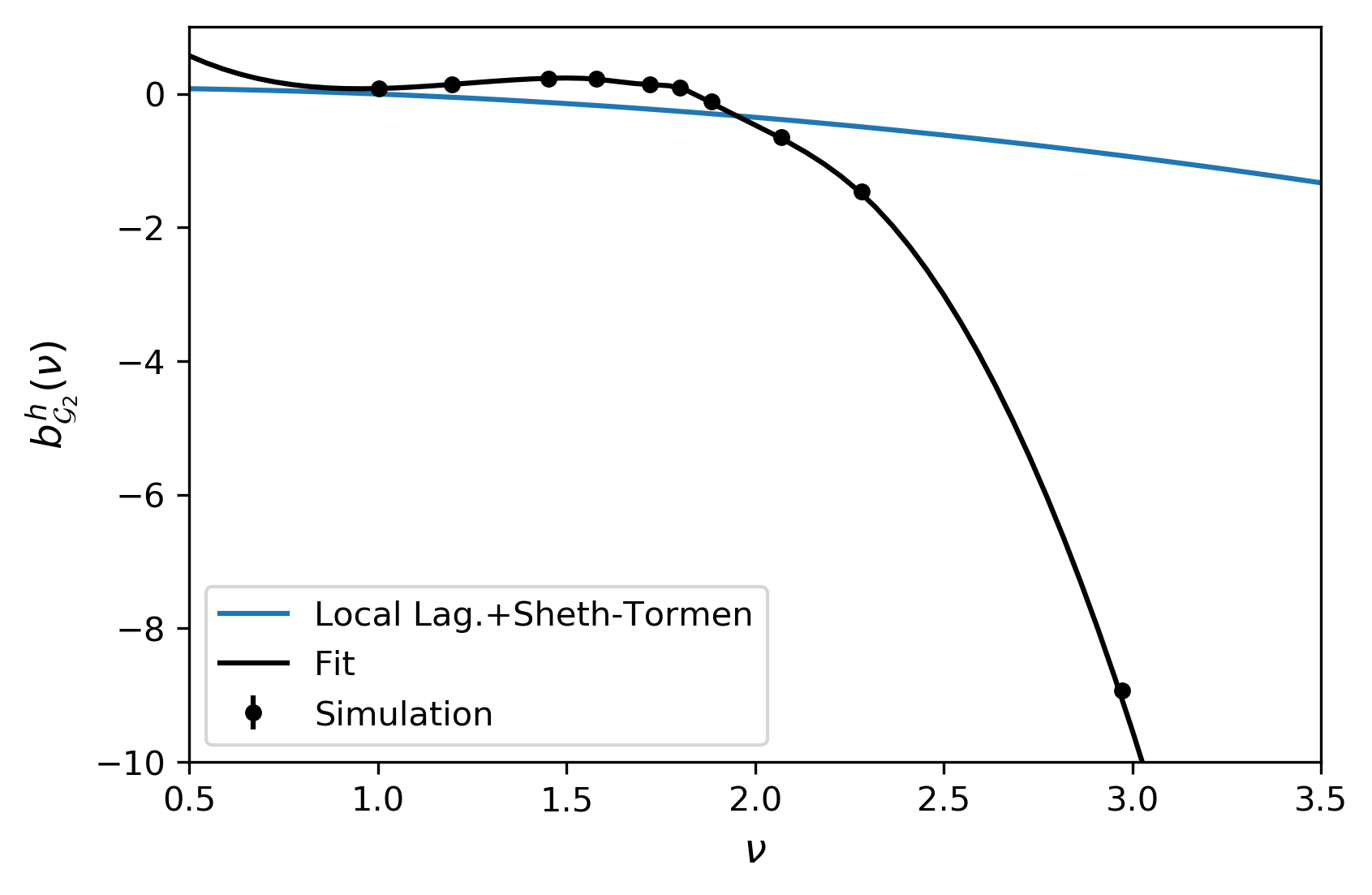}
\includegraphics[width=0.49\textwidth]{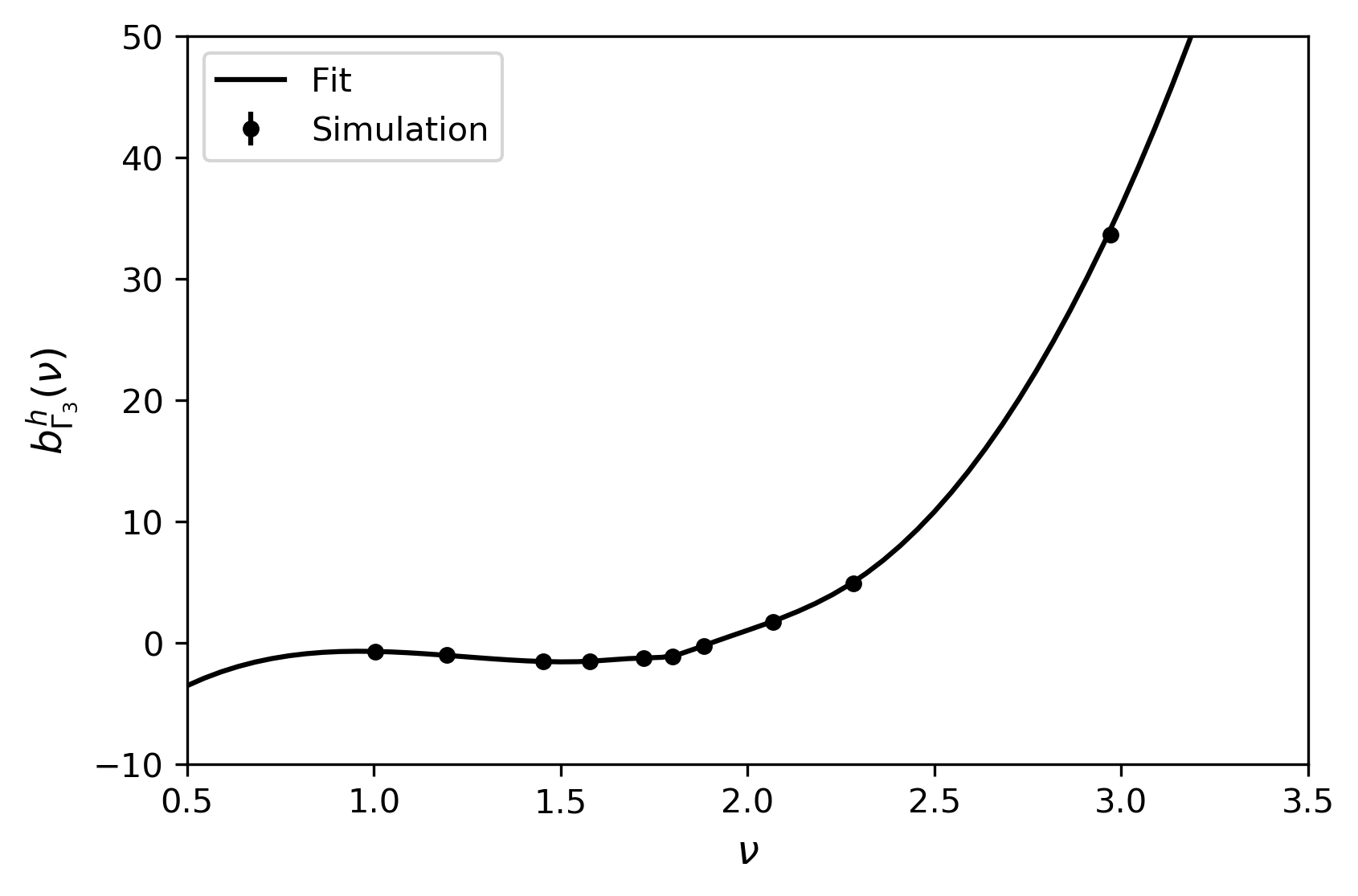}
\includegraphics[width=0.49\textwidth]{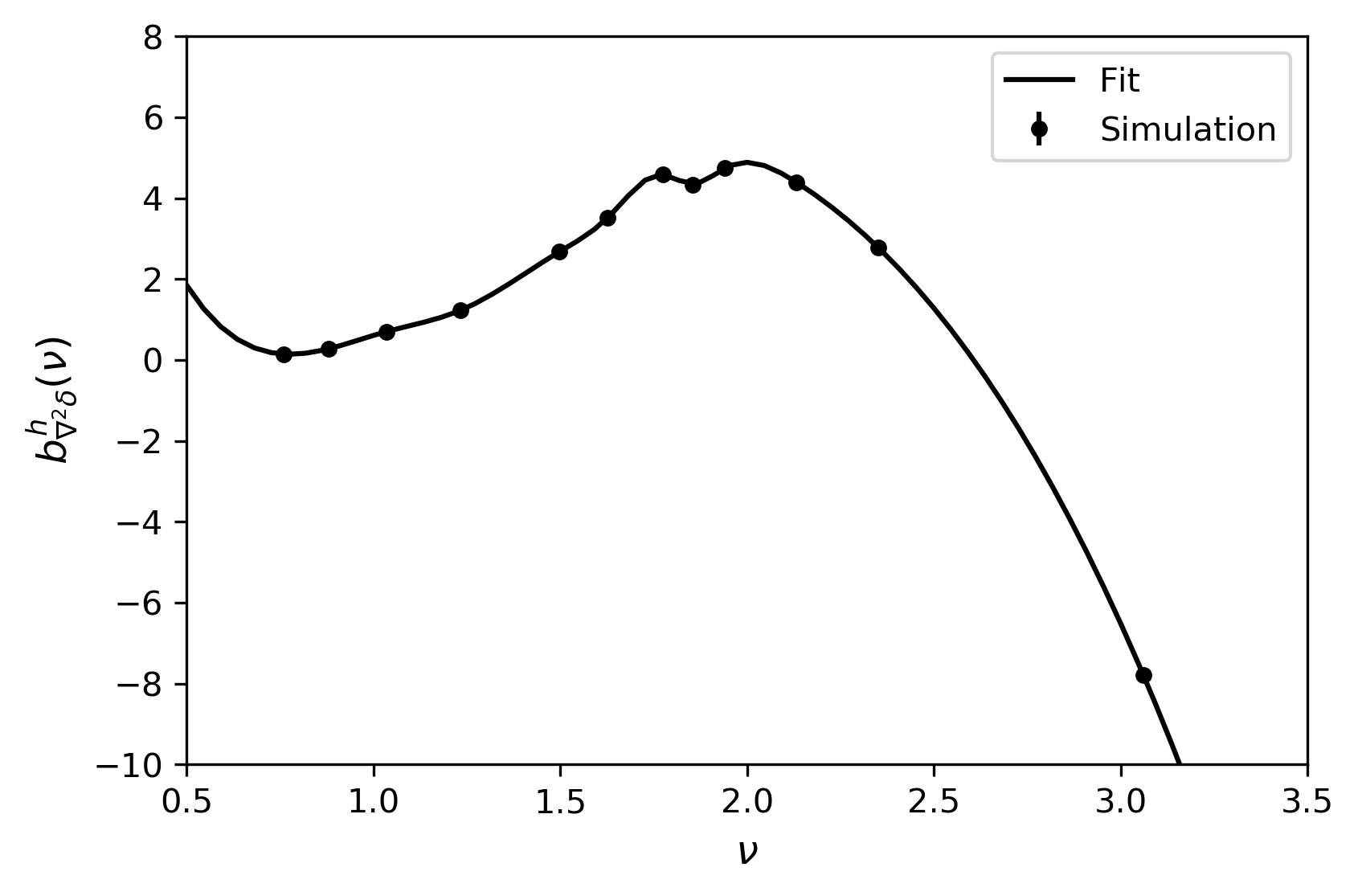}
   \caption{EFT parameters of halos as a function of the peak height $\nu$. 
   Some analytic models for $b_1(\nu)$
   and $b_{2}(\nu)$ are
   shown for comparison. Dots represent 
   measurements from N-body simulations,
   while the solid lines are 
   cubic spline fits. 
    } \label{fig:biasofnu}
\end{figure*}

In fig.~\ref{fig:biasofnu} we show the bias parameters of dark matter halos
as a function of $\nu$ extracted from
10 \texttt{Quijote} fiducial
simulations in ref.~\cite{Ivanov:2024xgb} at $z=0.5$.
We use the friends-of-friends halo catalogs. 
Since the cosmological parameters are fixed in this case, the variation of $\nu$
corresponds to the variation of the halo mass. 
As anticipated, 
all bias parameters exhibit clear correlations
with $\nu$. We use cubic splines 
to interpolate these dependencies. The interpolated 
functions $b_{\mathcal{O}_a}(\nu)$ 
($\mathcal{O}_a=\{\delta,\frac{1}{2}\delta^2,\G,\GG,b_{\nabla^2\delta}\}$) will be then used to 
express the bias parameters
as functions of mass.

For $b_1$ we also show the predictions
of the STHMF and THMF, 
while for $b_2$
we show the STHMF result~\eqref{eq:b2ST} where $b_{\G}$ is determined
by the local Lagrangian prediction.
We can see that for $b_1$
the agreement between the analytic formulas 
and measurements are quite decent, but 
as expected, it quickly deteriorates for $\nu>2$. 
As for $b_2$, the performance of the analytic 
prediction based on the Sheth-Tormen HMF is quite poor
for the entire range of masses.

\begin{figure*}
\centering
\includegraphics[width=0.49\textwidth]{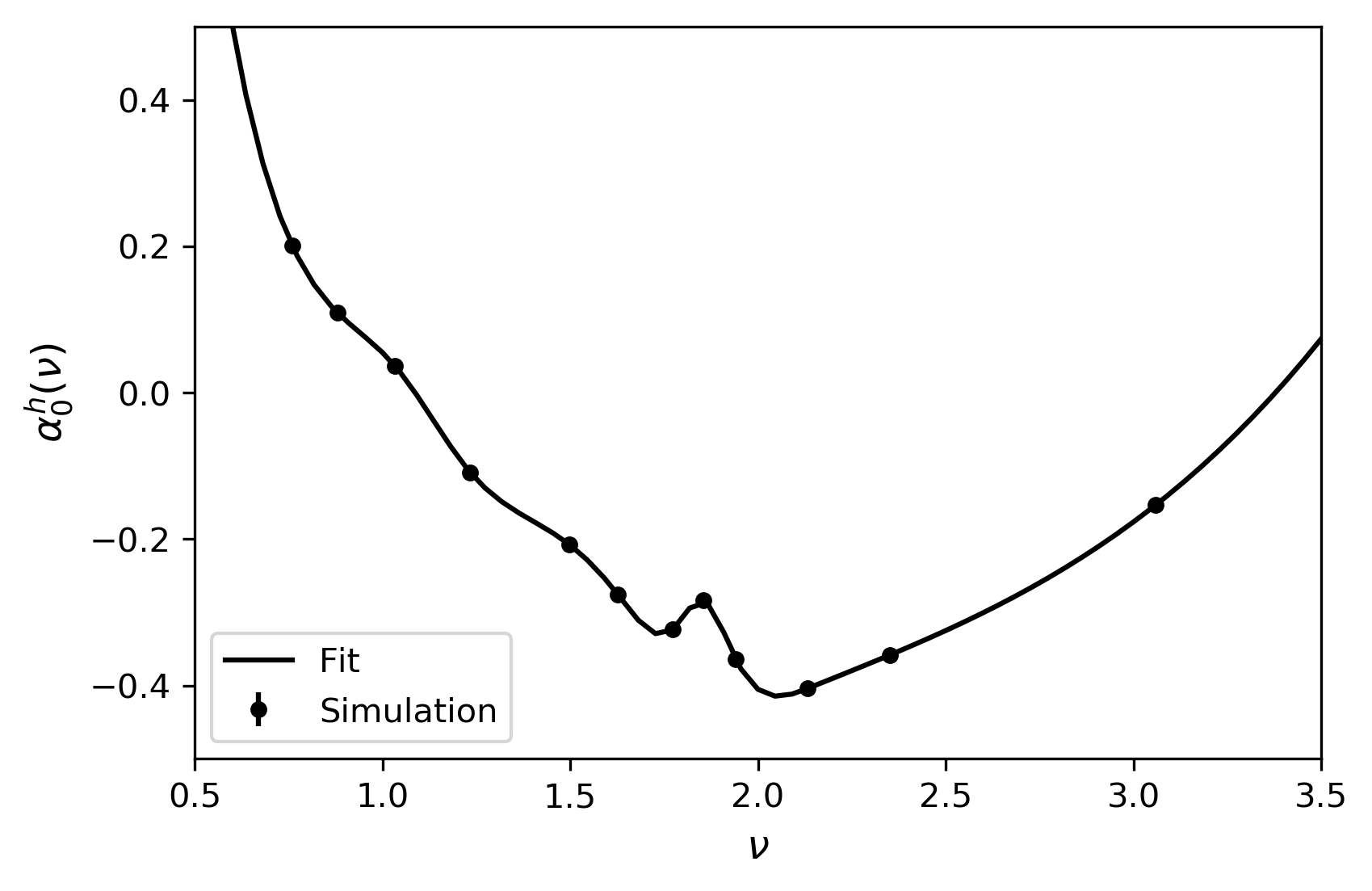}
\includegraphics[width=0.49\textwidth]{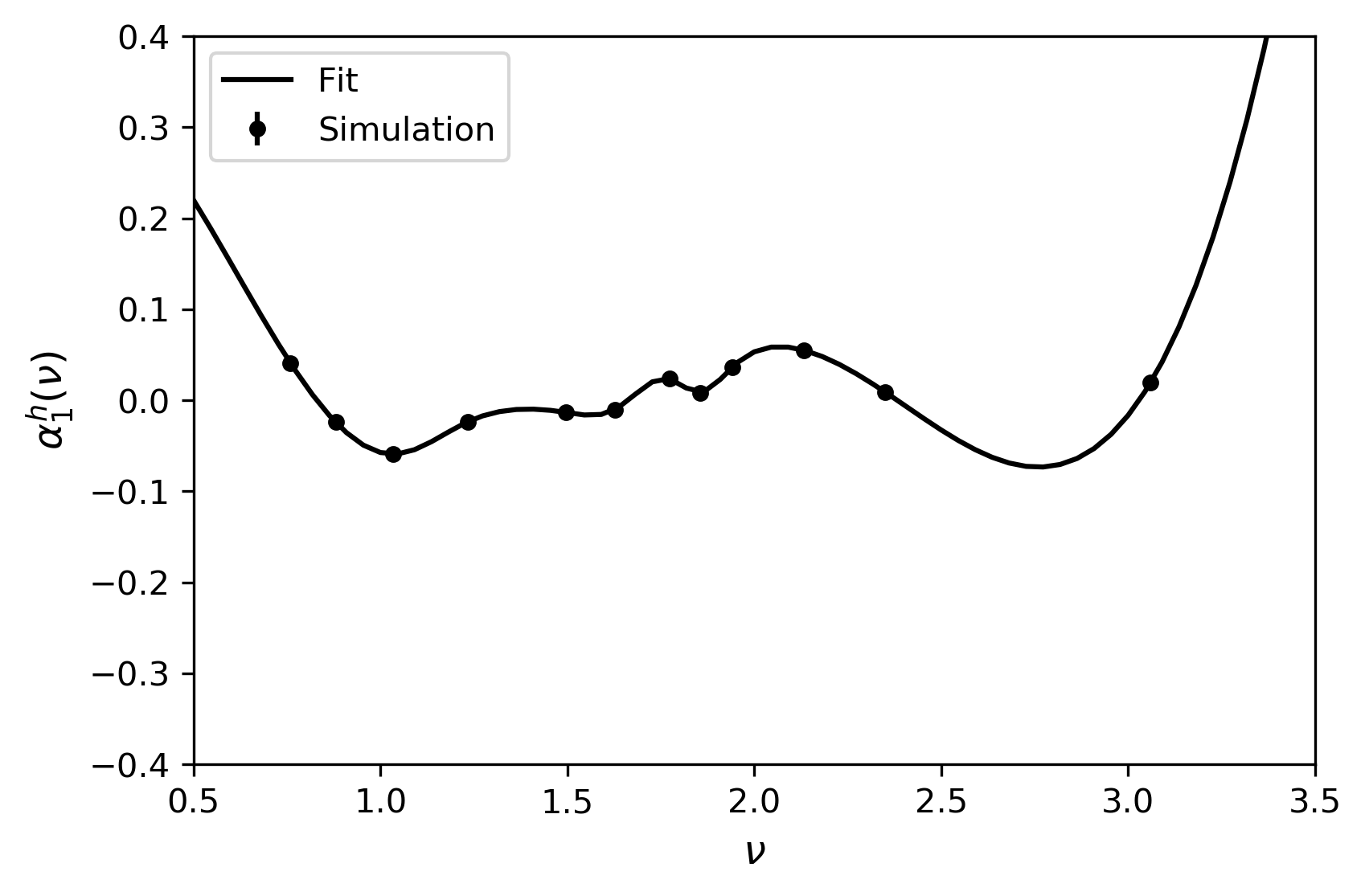}
\includegraphics[width=0.49\textwidth]{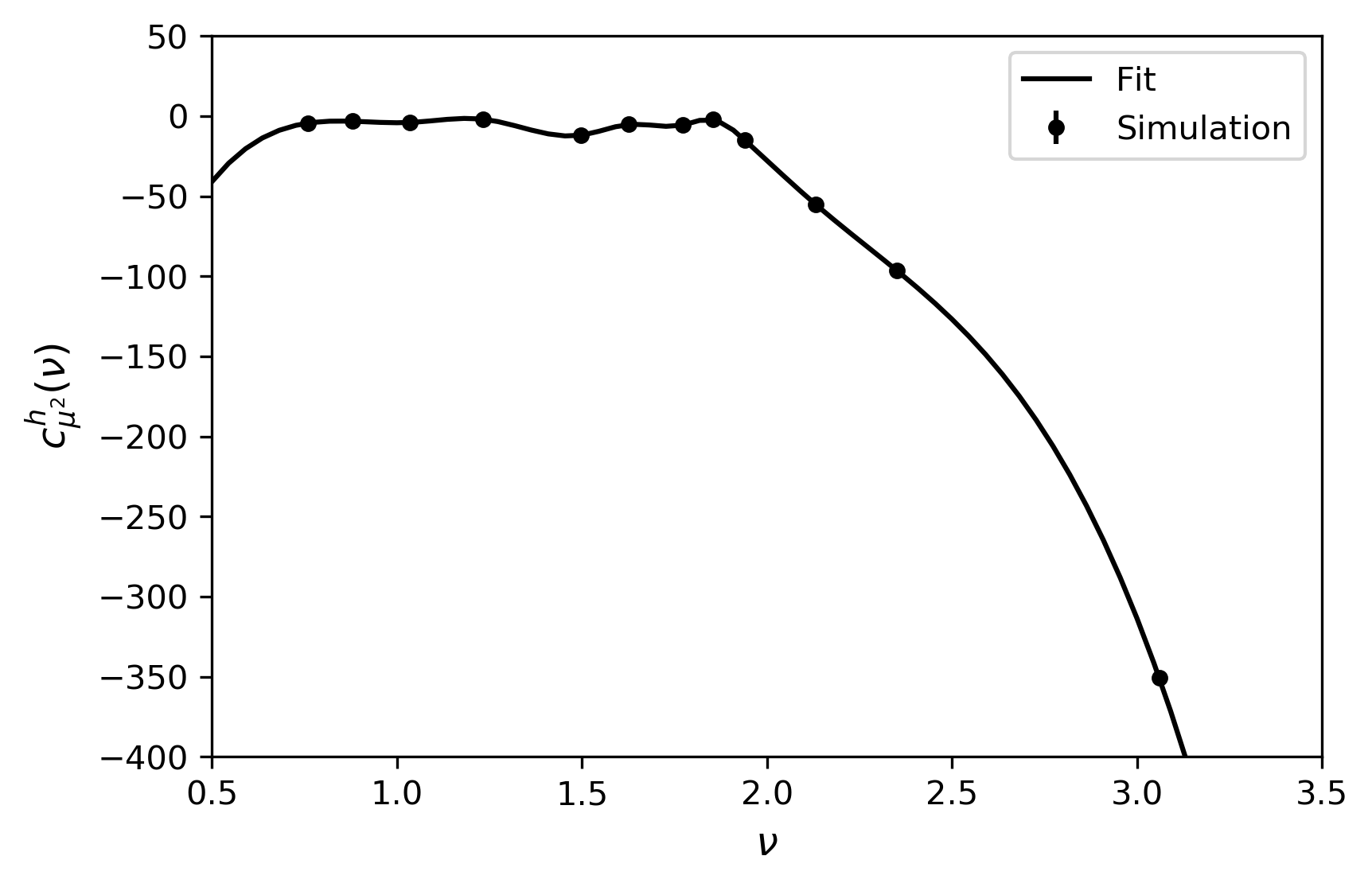}
\includegraphics[width=0.49\textwidth]{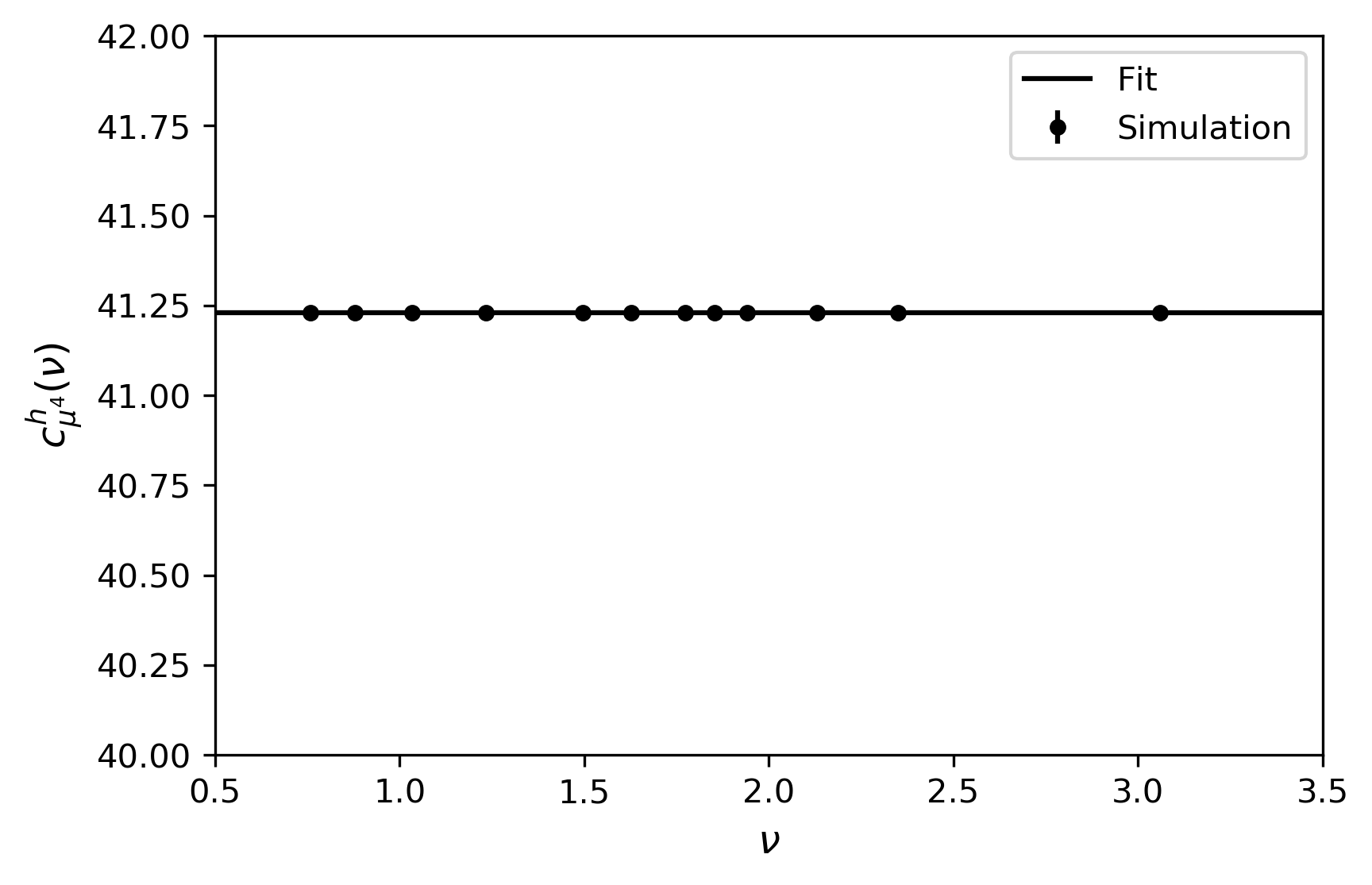}
\includegraphics[width=0.49\textwidth]{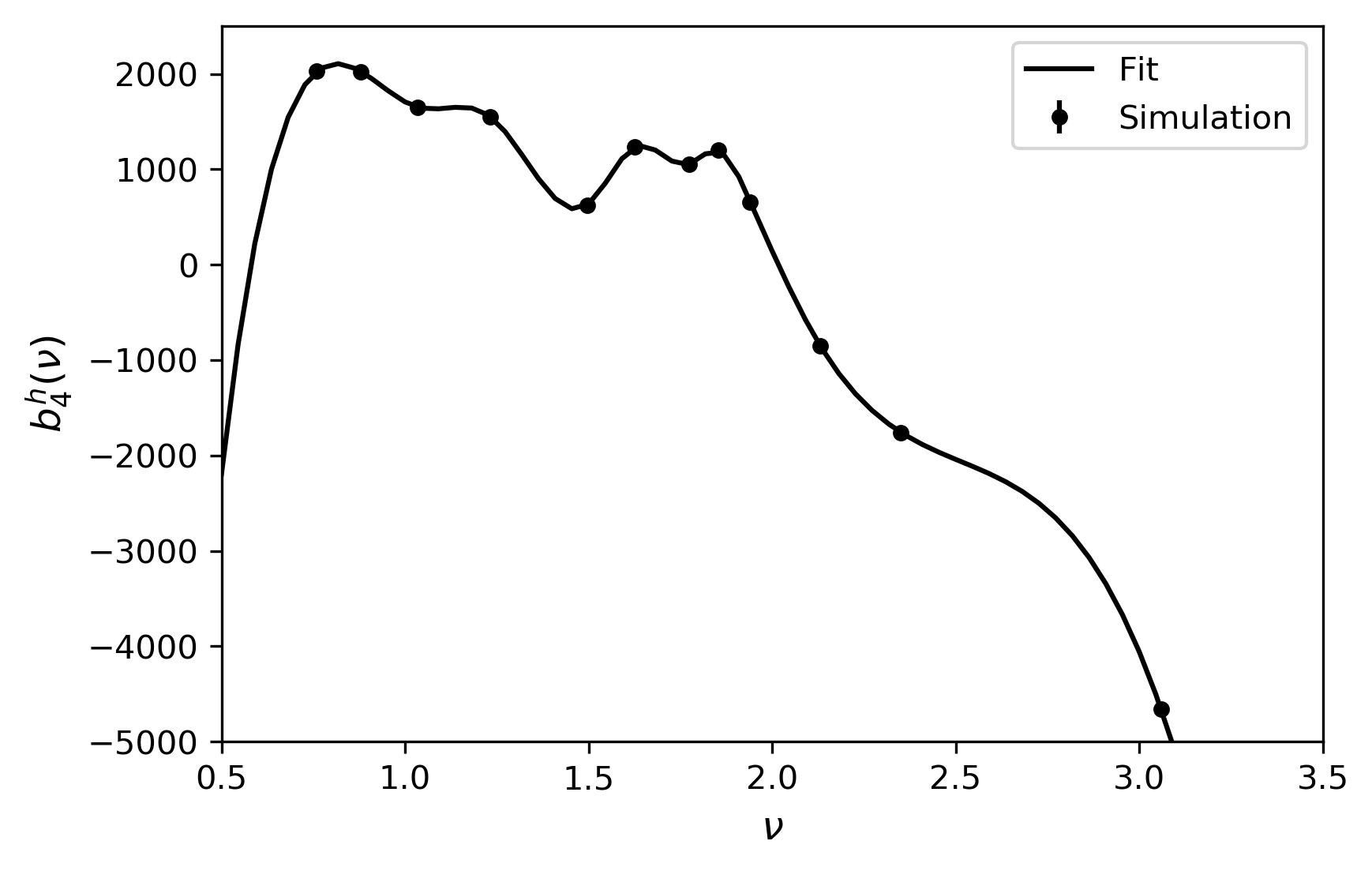}
\includegraphics[width=0.49\textwidth]{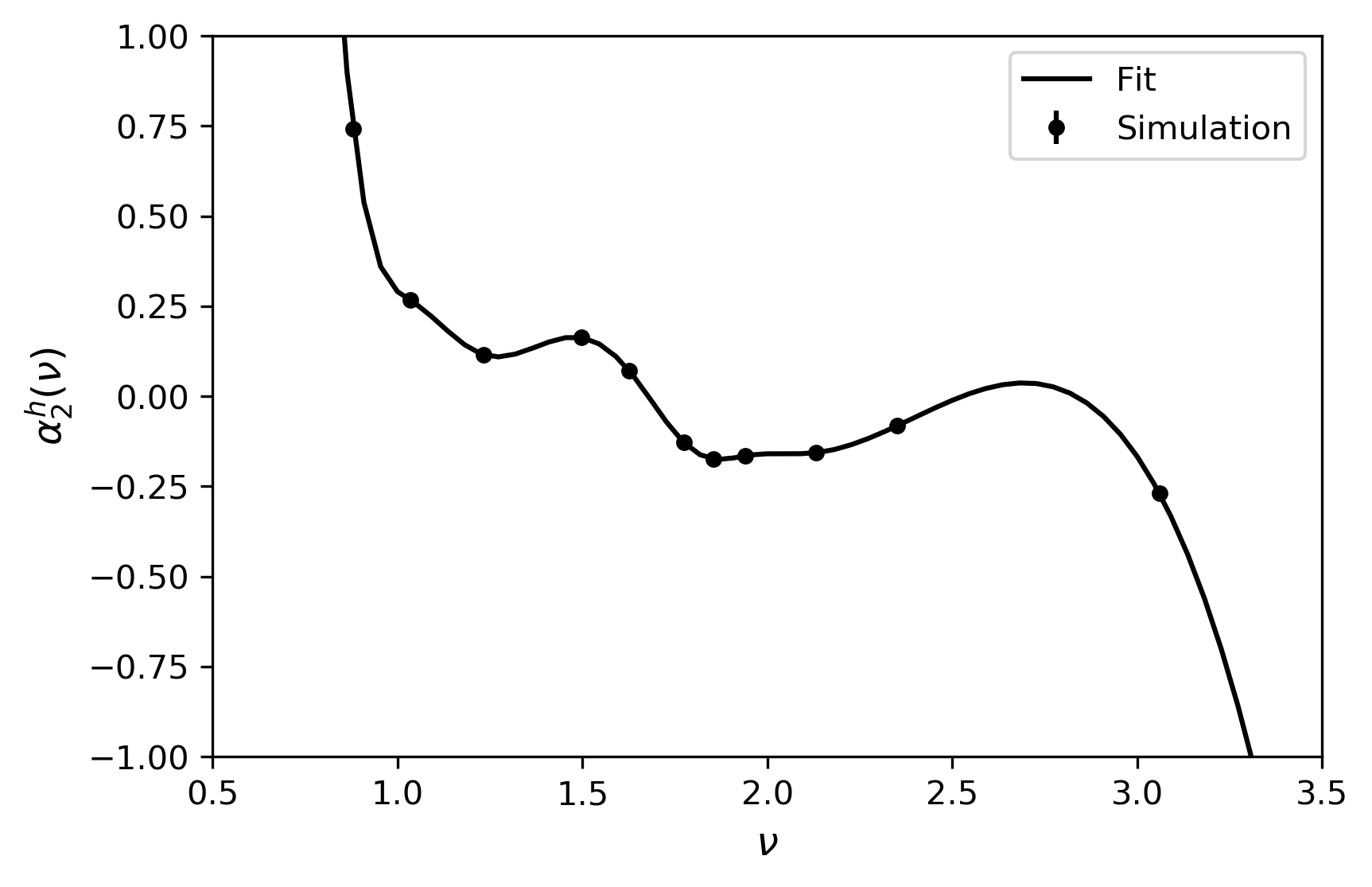}
   \caption{EFT stochastic parameters and  redshift-space counterterms of halos as a function of the peak height $\nu$. 
    } \label{fig:ctrofnu}
\end{figure*}

As far as the higher derivative counterterm
$b_{\nabla^2\delta}$ is concerned, 
we have previously found 
that its dependence on $\nu$
is halo-finder specific~\cite{Ivanov:2024xgb}.\footnote{Note that we have also found some halo-finder dependence for the usual bias parameters of halos as well~\cite{Ivanov:2024xgb}. 
However, this dependence is 
much weaker than that of the counterterms, 
and we stick
to the \texttt{Quijote} measurements for 
the bias parameters
in order to reduce the statistical noise
in the $b_{\mathcal{O}_a}(\nu)$
relations. 
}
In order to match our previous
HOD-based results from the \texttt{AbacusSummit} simulation more directly, 
we use the $b_{\nabla^2\delta}(\nu)$
measurements 
based on the same halo 
definition. 
Specifically, we measure $b_{\nabla^2\delta}(\nu)$
from a single box of \texttt{Abacus small} suite (site length of $500~\Mpch$)
for halo catalogs 
created with the CompaSO
halo finding algorithm~\cite{Hadzhiyska:2021zbd}. 
While the use
of the small \texttt{Abacus}
box leads to a somewhat large statistical
scatter for large values of $\nu$,
this affects our EFT parameter
predictions only at a $\sim 10\%$
level, which is within the target 
error budget.\footnote{In principle,
this can be improved by combining more
boxes or using the baseline 
\texttt{AbacusSummit} simulations. This is, however, not required for our current
precision goal. }
Note that one important feature of 
$b_{\nabla^2\delta}$ which we recover
both in \texttt{AbacusSummit} 
and \texttt{Quijote}
is its turnover at $\nu\approx 2$,
see the bottom right panel of fig.~\ref{fig:biasofnu}. 
This turnover is consistent 
with the previous measurements
by~\cite{Abidi:2018eyd}.

\subsubsection{Stochasticity parameters}

In the halo model, the stochastic 
contributions are captured by the one-halo term. 
In the first approximation, the stochastic power 
spectrum of halos can be approximated as Poissonian. 
In general, halos exhibit departures 
from the Poissonian stochasticity due to exclusion effects~\cite{Baldauf:2013hka}. 
These departures manifest themselves in the 
change of the overall amplitude of the constant contribution,
and the appearance of the scale-dependence. These are captured
within EFT by the following stochastic counterterms 
\be 
\begin{split}
\langle \varepsilon({\bf k})\varepsilon({\bf k}')\rangle = (2\pi)^3
\delta_D^{(3)}(\k+\k')P_{\rm stoch}\\
P_{\rm stoch}=\frac{1}{\bar n}\left(1+\alpha_0+\alpha_1\left(\frac{k}{\knl'}\right)^2\right)\,,
\end{split}
\ee 
where $\knl'=0.45~\hMpc$ following~\cite{Chudaykin:2020aoj,Chudaykin:2020hbf}.
Under our baseline assumption that the 
peak height is the only 
parameter that controls the stochasticity
of halos, 
we can extract the functions $\alpha_{0,1}(\nu)$
from the \texttt{Abacus small} CompaSO halo catalog, see the bottom panel of fig.~\ref{fig:ctrofnu}.
In agreement with the expected exclusion effects, we see that the low mass halos exhibit super-Poissonian stochasticity, while the large mass halos' 
stochasticity is sub-Poissonian~\cite{Baldauf:2013hka,Schmittfull:2018yuk}.
We also find that the scale-dependent stochasticity is negative 
for low mass halos, but it changes the sign and grows quickly 
for large halo masses.

\subsubsection{RSD counterterms}

In EFT, the RSD counterterms appear as a result
of the renormalization of local products of the density 
and velocity fields that stem from the redshift-space mapping. 
Expanding this mapping perturbatively we obtain 
the following expression for the halo density 
field in redshift space~\cite{Senatore:2014vja,Perko:2016puo}:  
\be 
\begin{split}
&\delta^{(s)}_h({\bm k})=\delta(\k)-\frac{ik_z}{\HH}v_{h,z}(\k)-
\frac{(ik_z)^2}{2\HH^2}[v^2_{h,z}]_\k\\
&-\frac{i^3}{3!}\left(
\frac{k_z}{\HH}\right)^3
[v^3_{h,z}]_\k - 
\frac{ik_z}{\HH}[v_{h,z}\delta_h]_\k
+\frac{(ik_z)^2}{2\HH^2}[\delta_h v_{h,z}]_\k~\,,
\end{split}
\ee 
where $\HH=aH$ and $v_h$ is the halo velocity field. Subscripts $z$ denote projections
onto the line-of-sight $z$. 
This renormalization can be thought of as a result of coarse-graining 
of the local products of the fields, and  practically 
it amounts to replacing the relevant products above as follows~\cite{Senatore:2014vja}
\be 
\label{eq:rsd_ctr}
\begin{split}
& \HH^{-1}[v_{h,z}\delta_h]_\k\to 
\HH^{-1}[v_{h,z}\delta_h]_\k 
+ic_4 k_z \delta_h \\
&\HH^{-2}[v^2_{h,z}]_\k\to 
\HH^{-2}[v^2_{h,z}]_\k
+c_2 \delta+c_3 \frac{k_z^2}{k^2}\delta_m,\\
& \HH^{-3}[v^3_{h,z}]_\k
\to \HH^{-3}[v^3_{h,z}]_\k+3 c_1 v_z\,,\\
& \HH^{-2}[v_z^2\delta_h]
\to \HH^{-2}[v_z^2\delta_h]
+c_5 \delta_h\,.
\end{split}
\ee 
Note that at leading order 
$v_h=v_{m}$, so that the counterterms
$c_{1,2,3}$ are the same 
for halos and dark matter thanks to the equivalence principle. 
For the linear velocity term in 
eq.~\eqref{eq:rsd_ctr}
we should also take into account 
its renormalization due to the 
dark matter velocity
counterterm
\be 
\theta =\theta^{(1)} + k^2 c_\theta \delta^{(1)}\,.
\ee 
In addition, there is a
halo-specific
velocity bias~\cite{Desjacques:2016bnm}
\be 
\theta =\theta^{(1)} + k^2 b_{\nabla^2{\bm v}} \delta^{(1)}\,,
\ee 
where the upper script $(1)$ means that
the corresponding fields should be 
evaluated in linear theory. 
Combining all contributions together, 
one obtains the following expression 
for the halo 
redshift space counterterm
\be 
\delta_{\rm ctr}^{(s)}=
\left(
c_{\mu^2}f\mu^2
+c_{\mu^4} f^2\mu^4
\right) k^2\delta^{(1)}\,,
\ee 
where  
\be 
\begin{split}
& fc_{\mu^2} \equiv (c_4 -\frac{1}{2}c_5)b_1 -\frac{c_2}{2}+fc_\theta + fb_{\nabla^2{\bf v}}\,,\\
& f^2c_{\mu^4}\equiv \frac{1}{2}(fc_1-c_3)\,.
\end{split}
\ee 
The $c_{\mu^4}$
counterterm is 
the same for halos and dark matter.
In contrast, 
the $c_{\mu^2}$
counterterm 
depends both on
cosmology and halo properties. 

Redshift-space clustering 
of matter and halos is subject to 
the non-linear smearing of power due 
to virial motions known as the 
finger-of-God effect~\cite{Jackson:2008yv}. In the EFT this leads to 
an enhancement of the EFT cutoff in redshift 
space and an early breakdown of the 
one-loop EFT model~\cite{Ivanov:2019pdj,Chudaykin:2020hbf,Ivanov:2021fbu}. One can improve 
the agreement between the one-loop
EFT calculation and data by including
the higher order counterterms in the calculation,
which capture the enhanced fingers-of-God.
Following refs.~\cite{Ivanov:2019pdj}, we parameterize them as 
\be 
\delta^{(s)}_{\delta_z^4\delta}=-\frac{b^h_4}{2}
f^4 \mu^4 k^4(b_1+f\mu^2)\delta^{(1)}\,,
\ee 
Finally, there is a stochastic
RSD counterterm parameterized
by $\alpha_2$:
\be 
P_{\rm stoch}=\frac{1}{\bar n}\left(1+\alpha_0+\alpha_1\left(\frac{k}{\knl'}\right)^2
+\alpha_2\mu^2 \left(\frac{k}{\knl'}\right)^2
\right)\,,
\ee 
where $\knl'=0.45~\hMpc$ as before. 

In fig.~\ref{fig:ctrofnu} we show the RSD counterterms $c_{\mu^2}$, $c_{\mu^4}$, $\alpha_2$
and $b_4$ as functions of $\nu$
extracted from the halo data.
These counterterms 
appear to be rather complicated, 
non-monotonic functions of $\nu$. 
Note that for the $c_{\mu^4}$ fit we have included
the dark matter field-level results as well
in order to reduce the scatter 
of measurements from large  
mass bins.

\subsection{HOD galaxies}

Within the HOD model, 
each halo can host galaxies
that follow specific probability distributions. 
In the simplest realization of these ideas,
the halo mass is the only parameter that determines
the number of galaxies.
The host galaxies are split into centrals and satellites. 
Each halo hosting a galaxy has exactly one central galaxy
which is located in the vicinity of the center of mass of the halo. 
All other galaxies within the same halo are 
called satellites. Each halo then is assigned a random 
variable of galaxies $N_g$, which is given by the sum of the 
central galaxy  $N_c$ (taking the values 0 or 1) 
and the satellite $N_s$, which is expressed as $N_s=N_c\mathcal{N}_s$
since the probability  to have a satellite galaxy is conditioned
on having a central.
The mean number 
density of galaxies is given by 
\be 
\bar n_g= \int d\ln M \bar n\langle N_g \rangle_M
\ee 
where $\langle N_g\rangle_M$ is the average halo occupation
distribution,
\be
\langle N_g \rangle_M = \langle N_c\rangle_M + \langle N_c\rangle_M\langle \mathcal{N}_s\rangle_M\,.
\ee
Note that we made an assumption that the HOD depends only on the halo mass. 
In particular, it is independent of the background cosmology. 
In addition, we will assume that $\mathcal{N}_s$ follows a 
Poisson distribution.

\subsubsection{Galaxy bias}

Perturbations in the halo number density
\be
n(M,\x)=\bar n(M)(1+\delta_h(\x))\,.
\ee
lead to perturbations in the galaxy density,
\be 
\begin{split}
& n_g(\x)=\bar n_g(1+\delta_g(\x))\\
&=\int d\ln M \bar n(M)\langle N_g\rangle_M(1+\delta_h(\x))~\,.
\end{split}
\ee
Removing the background quantities we find
\be 
\label{eq:dens_rel}
\begin{split}
& \delta_g(\x)=\frac{1}{\bar n_g}\int d\ln M \bar n(M)\langle N_g\rangle_M\delta_h(\x)~\,,
\end{split}
\ee
which implies the known relation 
between the bias parameters of halos and galaxies~\cite{Benson:1999mva,Akitsu:2024lyt}, 
\be 
\label{eq:biasHOD}
b^g_{\mathcal{O}_a}
=\frac{1}{\bar n_g}\int d\ln M \bar n(M)\langle N_g\rangle_M
b^h_{\mathcal{O}_a}(M)\,.
\ee 
Note that within the HOD model 
the above expression is often modified by the galaxy or halo
profile function $u(k)$. But the
perturbative bias expansion that we use is applicable
only on large scales, i.e. the limit $k\to 0$
where $u(k)\to 1$, which recovers eq.~\eqref{eq:biasHOD}.\footnote{In principle, 
the Taylor 
expansion of $u(k)$ produces $k^2$ corrections 
on large scales that modifies the higher derivative bias $b_{\nabla^2\delta}$.
In practice, however, we find that these corrections 
are negligibly small so we ignore them in what follows. }

The shape of $\langle N_g\rangle_M$
is modeled phenomenologically. In the simplest case, 
one assumes the following parameterization for the 
luminosity-limited galaxy samples, such as Luminous Red Galaxies~\cite{Zheng:2004id}, 
\be 
\label{eq:HODs}
\begin{split}
& \langle N_c\rangle =\frac{1}{2}\left[1+\text{Erf}\left(\frac{\log M - \log M_{\rm cut}}{\sqrt{2}\sigma}\right)\right]~\,,\\
& \langle \mathcal{N}_s\rangle = \Theta_H(M-\kappa M_{\rm cut})\left(\frac{M-\kappa M_{\rm cut}}{M_1}\right)^\alpha\,.
\end{split}
\ee 
This basic HOD model is parameterized by five free parameters
$M_{\rm cut}$, $\sigma$, $M_1$, $\kappa$ and $\alpha$
plus two velocity bias parameters 
that we will discuss shortly.

Note that there is an evidence
that the basic HOD model fails to 
correctly capture the clustering 
even for the luminosity-limited
galaxy samples, see e.g.~\cite{Hearin:2015jnf,Cuesta-Lazaro:2023gbv,Hahn:2023kky,SimBIG:2023nol,Hahn:2023udg,Hahn:2023udg}. HOD extensions with 
concentration and density-dependent assembly 
bias provide a better description
of the hydrodynamical simulations
and the data.
Such extensions are referred to as 
``decorated'' HOD models~\cite{Hearin:2015jnf}. 
The failure of the minimal HOD
is the main reason why realistic 
HOD-based EFT parameter samples applied to data 
have been
produced using the decorated HODs~\cite{Ivanov:2024hgq,Ivanov:2024xgb}. 
In what follows, however, we 
will use only this basic HOD parameterization
that allows for a simple analytic treatment.

\subsubsection{Redshift space counterterms and velocity bias}

Note that in redshift space, the density field in eq.~\eqref{eq:dens_rel}
is subjected to redshift-space mapping. Then one Taylor expands the mapping in the EFT
to arrive at the expressions of the type
(see eq.~\eqref{eq:rsd_ctr}):
\be
\label{eq:rsd_ctr_2}
 v^z_g(\x) \delta_g (\x)\big|_{\rm ren.} = v^z(\x) \delta_g(\x)
 -c_4^g \HH \nabla_z \delta^{(1)}_m(\x)\,,
\ee
where for convenience we switched to 
position space. 
Let us first assume that there is 
no velocity bias, i.e. ${\bm v}_g={\bm v}$, so that 
all galaxies are simply co-moving
with the halos. 
This should be satisfied on
large scales thanks to the 
equivalence principle. 
Then, in
the HOD model 
\be
\label{eq:hod_w_of_c4}
v_z \delta_g \Big|_{\rm ren.}  =  
\int dM \frac{d\bar n}{dM}
v_z \delta_h \Big|_{\rm ren.}
\,,
\ee
which
implies
\be 
\label{eq:ctr_weighting}
c_4^{g}=\frac{1}{\bar n_g}\int dM \frac{d\bar n}{dM}\langle
N_g\rangle c_4^h(M)~\,,
\ee 
i.e., all halo-dependent 
counterterms get weighted by the HOD.
We stress that so far we have ignored the velocity bias.
We conclude that the HOD framework predicts that  
the redshift-space
counterterms 
of galaxies 
is inherited 
from halos. 
Since the dark matter counterterms do not depend on the halo mass, we find 
\be 
c_{1,2,3}=
\frac{1}{\bar n_g}\int dM \frac{d\bar n}{dM}\langle
N_g\rangle c_{1,2,3}\,.
\ee 
Hence, they can be added 
to eq.~\eqref{eq:ctr_weighting}
so that we get 
the following expression for the total
redshift space counterterm
\be 
\label{eq:cgmu2}
c^g_{\mu^2}\Big|_{\rm halo} = 
\frac{1}{\bar n_g}\int dM \frac{d\bar n}{dM}\langle
N_g\rangle c^h_{\mu^2}(M)\,.
\ee 
The above equation
describes the contribution
to the RSD counterterm 
that galaxies ``inherit''
from the halos.

In addition to eq.~\eqref{eq:cgmu2},
there is  
a contribution
due to velocity bias 
w.r.t. dark matter particles 
within the halo,
which we denote as
$c^g_{\mu^2}\Big|_{\rm vel}$. 
Physically, they arise because
galaxies are moving w.r.t.
the halo center of mass
due to the virial potential 
within the halo. 
The basic halo models
describe such motions by means
of the Gaussian convolution 
of the density field in redsfhit space~\cite{Cooray:2002dia},
\be 
\label{eq:vel_disp}
\begin{split}
\delta^{(s)}_g &=(b_1+f\mu^2)\delta_m e^{-\frac{\Sigma^2_M\mu^2 k^2}{2}}\\
&=
(b_1+f\mu^2)\delta_m-\frac{k^2}{2}(b_1\Sigma^2_M \mu^2 + f\Sigma^2_M\mu^4)\delta_m
\\
&
+\frac{(k\mu\Sigma_M)^4}{8}(b_1+f\mu^2)\delta_m + ...
\end{split}
\ee 
The above Gaussian filter is determined by the 
mass-dependent
1-dimensional velocity
dispersion~\cite{Maus:2024dzi},
\be 
\Sigma_M =\left(\frac{GM}{2r_{\rm vir}}\right)^{1/2}\,.
\ee 
Eq.~\eqref{eq:vel_disp} 
describes the smearing 
of the density 
inside the halo. Without velocity bias, 
the centrals are co-moving with the 
halo centers of mass, so the smearing 
is produced only by the satellites. 
The $k^2\mu^2$ counterterm 
of galaxies then follows
from the HOD weighting of 
the above velocity dispersion~\cite{White:2000te,Seljak:2000jg,Maus:2024dzi}, 
\be 
\label{eq:hod_rsd_2}
\begin{split}
\delta^{(s)}_g&=\int dM \frac{d\bar n}{dM}\langle N_s \rangle \delta^{(s)}\\
&\supset 
-\frac{k^2\mu^2 }{2}\int dM \frac{d\bar n}{dM}\langle N_s \rangle b_1^h(M) \Sigma^2_M~\,.
\end{split}
\ee 
In the HOD framework
the 
velocity of 
galaxies is shifted from the 
velocity dark matter particles
by the velocity bias $\alpha_c$,
and $\alpha_s$ parameters,
respectively~\cite{Berlind:2001xk,Berlind:2002rn}, 
\be 
\begin{split}
{\bm v}_c = {\bm v}_{h}+\alpha_c ({\bm v}_{m}-{\bm v}_h)\,,\\
{\bm v}_s = {\bm v}_{h}+\alpha_s ({\bm v}_{m}-{\bm v}_h)\,,
\end{split}
\ee 
where 
${\bm v}_{h}$
is the halo center-of-mass 
velocity, 
and ${\bm v}_{m}$
is the velocity
of dark matter particle 
inside the halo. 
The case 
$\alpha_c=0$,  $\alpha_c=1$
reproduces the common  
assumption of the basic halo model
that the centrals are co-moving 
with the halo centers of mass, while
the satellites are moving with the typical
velocities of the host 
dark matter particles. 
Due to the velocity bias, 
the velocity dispersion
of galaxies is different 
from the matter velocity 
dispersion,
\be 
\Sigma^2_{c/s} = \alpha^2_{c/s}\Sigma^2_M\,.
\ee 
Note that due to the velocity bias the 
density component produced by the centrals
gets smeared as well. 
Including the velocity bias
and extending eq.~\eqref{eq:hod_rsd_2}
to the centrals 
we arrive at 
the final expression
\be 
\begin{split}
c^g_{\mu^2}\Big|_{\rm vel.}=&-\frac{1}{2f\bar n_g}\int dM 
\frac{d\bar n}{dM}b^h_1(M)\\
&
\times 
[\langle N_c\rangle 
\alpha^2_c + 
\langle N_s\rangle 
\alpha^2_s]
\Sigma^2_M\,.
\end{split}
\ee 
The total $\mu^2k^2\delta_m$
counterterm coefficient is the sum 
of the halo one and the 
one produced 
by the virialized velocity 
dispersion, 
\be 
c_{\mu^2}=c_{\mu^2}\Big|_{\rm halo}
+c_{\mu^2}\Big|_{\rm vel.}
\,.
\ee 

In passing, 
we note that the halo model 
expression~\eqref{eq:vel_disp}
predicts that the coefficient
in front of $k^2\mu^4$ term is 
mass dependent, 
\be 
c_{\mu^4} = -\frac{1}{2f}\Sigma^2_M\,.
\ee 
The non-universality of this term 
clashes 
with the EFT prediction dictated by the equivalence
principle. This is another limitation
of the halo model, which appears to 
violate the equivalence principle 
in addition to the mass and momentum conservation.

Let us discuss now 
the higher-order $k^4$  counterterm 
parameterized by the constant 
$b_4$. 
Just like the $\mu^2 k^2$
counterterm, the $\mu^4 k^4$ counterterm
also consists of two pieces,
\be 
b^g_4 = b^g_4\Big|_{\rm halo} + b^g_4\Big|_{\rm vel.}
\ee 
The first 
piece is inherited from the 
dark matter halos, 
\be 
b_1^g b^g_4\Big|_{\rm halo} =\frac{1}{\bar n_g}\int dM \frac{d\bar n}{dM}\langle N_g \rangle b^h_1(M)b^h_4(M)~\,.
\ee 
The second term $b^g_4\Big|_{\rm vel.}$
is generated by the velocity bias, and can be obtained
by Taylor expanding 
the Gaussian smearing kernel 
in eq.~\eqref{eq:vel_disp}
up to quadratic order. 
We note, however, that at this order
the phenomenological models
available in the literature start 
to give different predictions. 
The Gaussian model gives a coefficient 
$1/8$ in front of the $(k\mu\Sigma_M)^4$
term, but the other popular
Lorentzian model of the fingers-of-God
damping~\cite{Hand:2017ilm} gives the coefficient $3/16$. More generally, 
one expects a more complicated 
fingers-of-God damping function 
modified by the higher order moments 
of the virialized velocity field~\cite{Scoccimarro:2004tg,eBOSS:2018whb,Eggemeier:2025xwi}. In this work we stick 
to the simplest model where the 
only relevant parameter
is the velocity dispersion, 
and use our flexibility 
in the choice of the coefficient
in front of the $(k\mu\Sigma_M)^4$
to set it to $1/16$, 
preferred by the field-level
transfer function data 
from~\cite{Ivanov:2024xgb}. 
Taking this model 
and rescaling the 
velocity dispersion by the appropriate 
velocity bias
coefficients, we obtain
\be 
\begin{split}
 b^g_{4}\Big|_{\rm vel.}=&
-\frac{1}{8f^4  b_1^g \bar n_g}\int dM 
\frac{d\bar n}{dM} \Sigma^4_M b_1^h(M)\\
& \times [\langle N_c\rangle 
\alpha^4_c + 
\langle N_s\rangle 
\alpha^4_s]\,.
\end{split}
\ee 
It is interesting to note
that the net correction to $b_4$
from the virialized motion is
negative, which can be contrasted
with  $b^g_4\Big|_{\rm halo}$,
which is positive for relatively light
halos and negative for 
very massive ones, see fig.~\ref{fig:ctrofnu}.
Thus, the total sign of the $b_4^g$
term can be both positive and negative
depending on the HOD model. 

Our results clarify the role of the satellites in limiting the range of validity
of EFT in redshift space. 
The latter is determined by the 
amplitude of the EFT redshift space parameters.
In the case of the basic HOD model
(i.e. if we ignore the counterterm
contributions inherited from halo), 
$\alpha_c=0$ and $\alpha_s=1$,
i.e. the velocity dispersion of the 
satellites is the only source of the 
fingers-of-God.  
However, analyses 
of the galaxy samples
with reduced satellite fractions~\cite{Ivanov:2021zmi,BaleatoLizancos:2025wdg}
showed that their removal does 
not significantly improve 
the reach of the one-loop
EFT power spectrum predictions
in redshift space. 
Our analysis shows that one reason
for that is the RSD counterterm 
contribution
inherited from the halos, eq.~\eqref{eq:cgmu2}. This contribution
can be quite large 
even for the HODs with centrals only, 
especially if one is tracing 
massive halos.

\subsubsection{Galaxy stochasticity}

In the basic formulation of the halo model
the stochasticity of halos and galaxies 
is captured by the 
correlations between galaxies
within the same halo, encapsulated by
one-halo term. 
Clearly, by construction 
this term can produce only the 
super-Poisson contribution
~\cite{Abramo:2015daa},
\be 
\label{eq:Pg_sat}
\begin{split}
\Delta P^g_{\rm stoch.}&=\frac{1}{\bar n_g^2}\int d\ln M \bar n \langle N_c\rangle (2 \langle \mathcal{N}_s\rangle +\langle \mathcal{N}_s\rangle^2 )\,.
\end{split}
\ee 
The stochasticity of
the halos is assumed to be Poissonian and given by $1/\bar n$. Likewise, 
the stochasticiy of the centrals
is also given by the Poissonian
prediction. In the halo model terminology
this is a stochastic part of the two-halo term. 
Combined with 
the two-halo satellite Poissonian prediction, this 
yields the standard answer 
$\bar{n}_g^{-1}$. 
Combining this with the one-halo satellite 
contribution~\eqref{eq:Pg_sat}, 
we get
\be 
\label{eq:Pg_simple}
\begin{split}
P^g_{\rm stoch}&=\frac{1}{\bar n_g}+\frac{1}{\bar n_g^2}\int d\ln M \bar n_h\langle N_c\rangle (2 \langle \mathcal{N}_s\rangle +\langle \mathcal{N}_s\rangle^2 )\,,\\
&=\frac{1}{\bar n_g^2}\int d\ln M \bar n_h\langle N_c\rangle (1+3 \langle \mathcal{N}_s\rangle +\langle \mathcal{N}_s\rangle^2)\,.
\end{split}
\ee 
While this  equation does 
capture the basic fact that the galaxy stochasticity
is different from the Poisson prediction, 
it misses the exclusion effects that produce
the sub-Poissonian shot noise~\cite{Baldauf:2013hka}. 

The proper inclusion of exclusion effects 
requires taking into account deterministic 
two-halo correlations. 
In ref.~\cite{Baldauf:2013hka} this was done 
for the galaxy bias model including the Lagrangian 
quadratic bias. 
In principle, one could extend the approach of.~\cite{Baldauf:2013hka}
to higher order biases~\cite{Hand:2017ilm,Sullivan:2021sof}. 
But even in this case the exclusion prediction
depends on the galaxy correlation function
on very small scales, which cannot be predicted analytically.  
Therefore, for the phenomenological reasons, we consider 
a simple modification of~eq.~\eqref{eq:Pg_simple}
with the one-halo term replaced by its 
exclusion-corrected version from simulations,
\be 
\label{eq:Pg_naive}
\begin{split}
&P^g_{\rm stoch}=\frac{1}{\bar n_g^2}\int d\ln M \bar n \big[\langle N_c\rangle \\
&\times (1+3 \langle \mathcal{N}_s\rangle+\langle \mathcal{N}_s\rangle^2 )(1+\alpha_0(\nu[M]))\big]\,,
\end{split}
\ee 
implying 
\be 
\label{eq:a0_HOD}
\begin{split}
\tilde{\alpha}_0^g = &\frac{1}{\bar n_g}\int d\ln M \bar n\big[ \langle N_g\rangle \alpha_0(M)\\
&+
(1+\alpha_0(M))(2\langle N_c\rangle\langle \mathcal{N}_s\rangle+\langle N_c\rangle\langle \mathcal{N}_s\rangle^2 )\big]\,.
\end{split}
\ee 
This phenomenological model propagates the halo exclusion 
to galaxies, but it still does 
not correctly reproduce the stochasticity parameters measured 
from simulations: it over-predicts the 
super-Possonian stochasticity by a factor of $10$
and under-predicts the sub-Poissonian behavior. 
To compensate for the former, we multiply the answer by a fudge factor $g_{\alpha_0}=0.072$.
To account for the latter, we add the exclusion window contribution, which
we weight over the HOD and correct by another fudge factor $g_{\rm excl}=0.2~[\hMpc]^3$, resulting in 
\be 
\alpha_0^g =g_{\alpha_0}\tilde{\alpha}_0^g -\frac{4\pi g_{\rm excl}}{3\bar n_g}\int d\ln M \bar n_h r^3_{\rm vir}(M)
\langle N_g\rangle~\,,
\ee 
where $r_{\rm vir}=(3M/(800\bar \rho_m))^{1/3}$.
The resulting model reproduces the distribution of stochasticity 
EFT parameters quite well.

As far as the scale dependent stochasticity 
parameter is concerned, eq.~\eqref{eq:Pg_simple}
suggests the following phenomenological
modification, 
\be 
\label{eq:a1_naive}
\begin{split}
&\tilde{\alpha}^g_1=\frac{1}{\bar n_g}\int d\ln M \bar n~\alpha_1(M)\big[\langle N_c\rangle  (1+3 \langle \mathcal{N}_s\rangle+\langle \mathcal{N}_s\rangle^2 )\big]\,.
\end{split}
\ee 
To partly account for the exclusion effects, we add a weighted contribution
from the hard sphere halo model
of~\cite{Baldauf:2013hka}
and introduce two additional 
fudge factors which we calibrate 
from simulations, 
\be 
\label{eq:a1_final}
\begin{split}
&\alpha^g_1=g_{\tilde{\alpha}_1}\tilde{\alpha}^g_1 + \frac{4\pi g_{\alpha_1}}{30\bar n_g}
\int d\ln M \bar n r^5_{\rm vir}(M)
\langle N_g\rangle
\end{split}
\ee 
In practice, we use
\be 
g_{\tilde{\alpha}_1}=5\,,\quad g_{\alpha_1}=2~[\hMpc]^5\,.
\ee 
We adopt the above model 
as a baseline in what follows.

Finally, the halo model can be used to 
compute the stochastic $k^2\mu^2$
counterterm in redshift space. As suggested
by our previous discussion on RSD counterterms, we split 
the total counterterm into two pieces: the one 
``inherited'' from halos ($\alpha_2^g \Big|_{\rm halo}$), 
and the one generated by the velocity bias
($\alpha_2^g \Big|_{\rm vel.}$), 
for which we use expressions 
involving the velocity dispersion. 
The latter corresponds 
to the 
redshift-space modification of the 
one-halo term and it 
has been studied 
in detail
in the literature. It has a structure
very similar to that of the 
super-Poissonian shot noise~\cite{Seljak:2000jg,Maus:2024dzi}, 
\be 
\begin{split}
& \alpha_2^g = -\frac{1}{2\bar n_g^2}
\int d\ln M \bar n
\Sigma^2_M
\langle N_c\rangle 
(2\langle \mathcal{N}_s\rangle+
2\langle \mathcal{N}_s\rangle^2
)\,,
\end{split}
\ee 
where an additional factor of 2
in front of $\langle \mathcal{N}_s\rangle^2$ above appeared
because both satellite densities 
inside the halo is smeared
by the virial motion. 
The problem with the above expression
is that it does not correctly
capture the velocity bias
of the centrals. One possible 
resolution of this issue  
is to add the smeared combination
of the one-halo term and the 
shot noise contribution, resulting 
in the expression, 
\be 
\begin{split}
& -\frac{1}{2\bar n_g^2}
\int d\ln M \bar n
\Sigma^2_M
\langle N_c\rangle 
(1+3\langle \mathcal{N}_s\rangle+
2\langle \mathcal{N}_s\rangle^2
)\,. 
\end{split}
\ee 
This expression is phenomenological,
and thus we anticipate that it will 
require fudge factors. 
Applying the velocity bias
to the satellites
and the centrals in the above expression, 
and introducing fudge factors $g_{\alpha_c},g_{\alpha_s}$
we finally get 
\be 
\label{eq:rsd_a2_fudge}
\begin{split}
\alpha_2^g \Big|_{\rm vel.}= & 
-\int d\ln M \bar n \Sigma^2_M
\langle N_c\rangle 
\\
&\times [ g_{\alpha_c}
\alpha_c^2
+g_{\alpha_s}\alpha_s^2
\langle \mathcal{N}_s\rangle
(3 
+ \langle \mathcal{N}_s\rangle) ]\,.
\end{split} 
\ee  
The final expression
for $\alpha_2^g$ is obtained
by adding the 
piece inherited 
from the halo centers of mass, 
\be 
\label{eq:rsd_a2_fudge}
\begin{split}
\alpha_2^g \Big|_{\rm halo}= & 
\int d\ln M \bar n 
\alpha_2^h(\nu)
\langle N_c\rangle 
 [ 1+
\langle \mathcal{N}_s\rangle
(3 
+ \langle \mathcal{N}_s\rangle) ]\,.
\end{split} 
\ee 
Putting all together, 
we write down the final model
\be 
\label{eq:final_a2g}
\alpha_2^g =g_{\alpha_2^h}\alpha_2^g \Big|_{\rm halo}+\alpha_2^g \Big|_{\rm vel.}\,,
\ee 
where $\alpha_2^g$ is another fudge factor. 
A good match to data is obtained by setting
\be 
g_{\alpha_2^h} = 0.2\,,\quad 
g_{\alpha_c}= -1.5 \,,
\quad 
g_{\alpha_s}= 0.0033~\,,
\ee 
where the last two factors are given in units of $[\hMpc]^2$.
Comparing these fudge factors
with the predictions of the naive halo model, we 
see that the numerical data 
actually prefers a strong positive 
contribution
from the centrals due to their 
velocity dispersion, 
a somewhat weaker 
contribution
from the halo centers of mass, 
and a noticeably weaker
contributions from the satellites~\cite{Ivanov:2024jtl}. 
The fact that the 1-halo term 
overprotects the large and negative 
$\alpha^g_2$, which 
contradicts the N-body data was 
previously pointed out by~\cite{Maus:2024dzi}.

It is suggestive that 
the tension between the halo model
predictions for the stochastic RSD
counterterm 
could be resolved once 
the halo exclusion effects are
taken into account. 
In addition, the positive sign of 
$\alpha_2$ obtained in N-body simulations
could be explained by the
modulation of the
galaxy power spectrum by the 
correlation function
of virialized velocity 
fields inside the halo~\cite{Maus:2024dzi}.
We defer 
a detailed analytic modeling of $\alpha_2$
to future work,
and use the phenomenological
expression~\eqref{eq:final_a2g}
in the rest of our paper.



\begin{figure*}[ht!]
\centering
\includegraphics[width=1.00\textwidth]{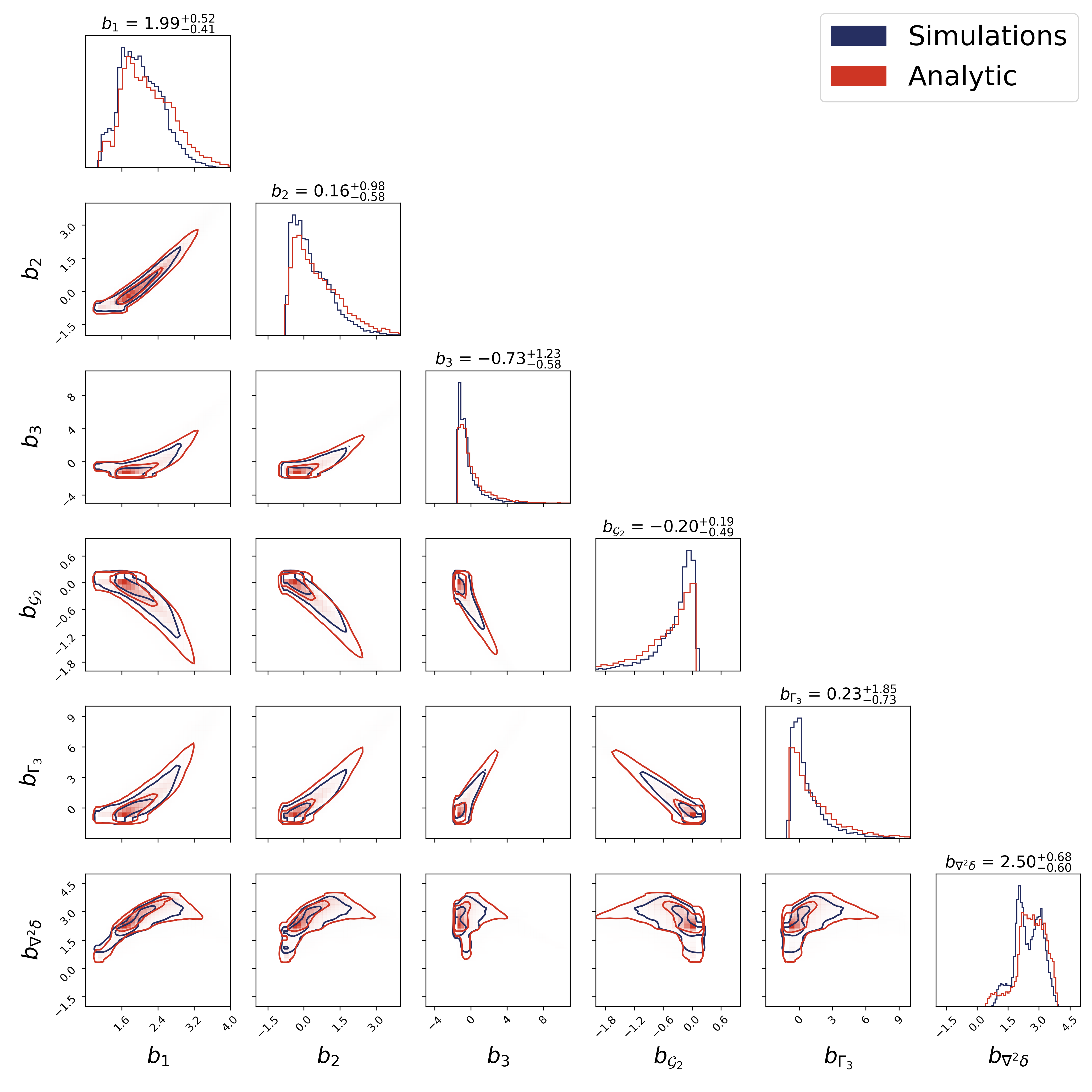}
   \caption{
The density of EFT galaxy bias parameters
for HOD models extracted from 
N-body simulations (blue) and 
generated with our analytic model (red). 
Density levels correspond to two-dimensional $1$-$\sigma$
  and $2$-$\sigma$ intervals (i.e. 39.3\% and 86.5\% of samples). 
    } \label{fig:dist_QH}
\end{figure*}

\begin{figure*}[ht!]
\centering
\includegraphics[width=1.00\textwidth]{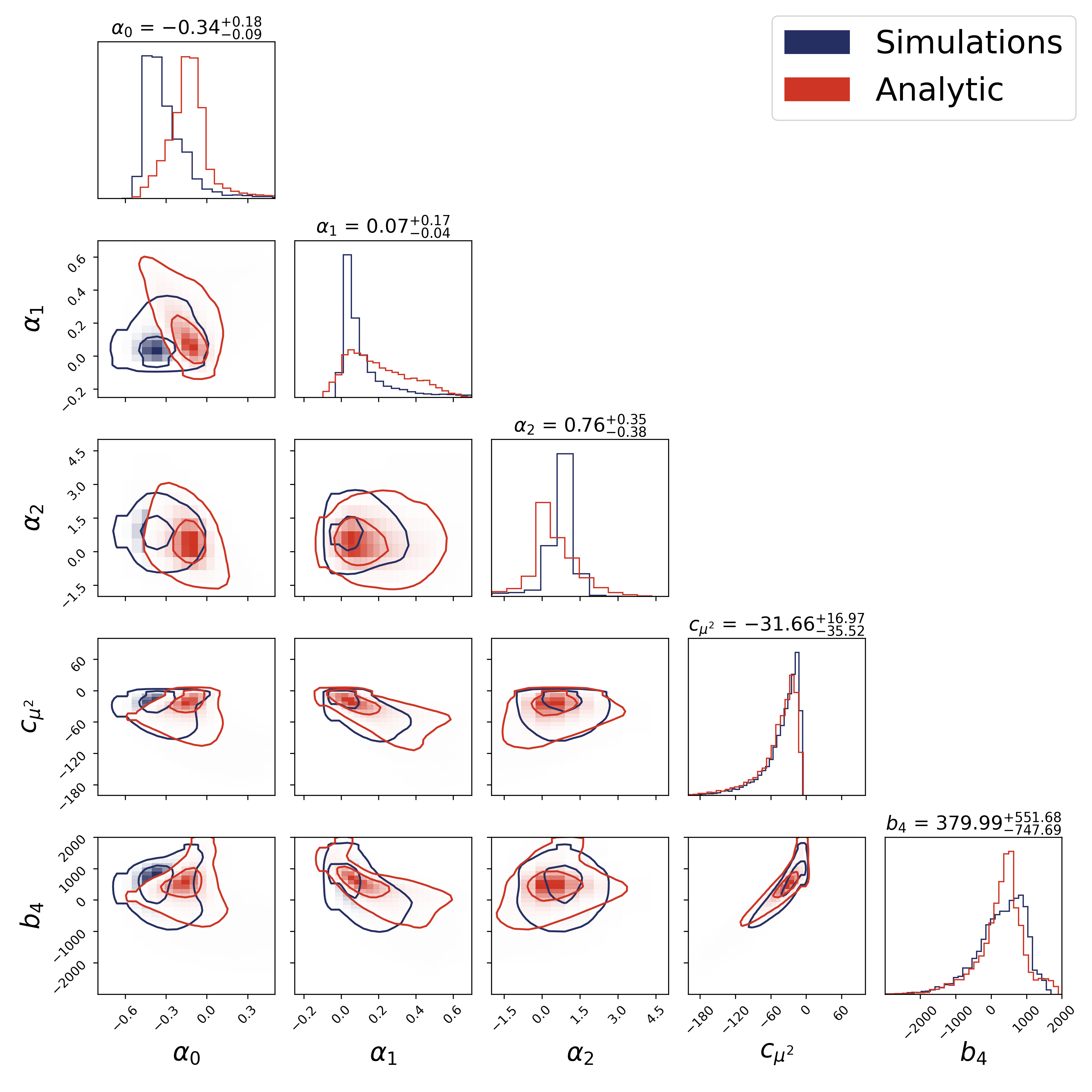}
   \caption{Same as fig.~\ref{fig:dist_QH}
   but for the stochastic 
   EFT parameters and redshift-space
   counterterms. 
    } \label{fig:dist_QH_ctr}
\end{figure*}

\section{Comparison with simulations}
\label{sec:sim}

In this section we
compare our 
analytic predictions
with the EFT parameter
measurements 
from \texttt{AbacusSummit} simulations~\cite{Maksimova:2021ynf}
carried out in~\cite{Ivanov:2024hgq,Ivanov:2024xgb}.The caveat here is that these works
produced EFT samples for decorated HODs.
In order to produce the samples 
for the base HOD we use 
the EFT parameter distribution
conditioned on the HOD parameters
estimated by the normalizing flows
following the 
approach of ref.~\cite{Ivanov:2024hgq}.

We generate the samples
of EFT parameters
by numerically 
evaluating 
our analytical formulas 
for a large grid 
of HOD parameters, 
which we randomly sample
from the following 
flat distribution
from~\cite{Ivanov:2024hgq}:
\be 
\label{eq:lrghod1}
\begin{split}
& \log_{10} M_{\rm cut}\in [12.4,13.3]\,,\quad 
 \log_{10}  M_1\in [13.2,14.4]\,,\\
& \log_{10} \sigma \in [-3.0,0.0]\,,\quad  \alpha \in [0.7,1.5]\,,\quad \kappa \in [0.0,1.5]\,,\\
&\alpha_c \in [0,0.5]\,,\quad 
\alpha_s \in [0.7,1.3]~\,,
\end{split}
\ee 
where the halo masses are given
in units of $h^{-1}M_\odot$.
With a slight abuse of terminology, 
we will call the HOD priors generated by 
numerically evaluating our analytic expressions
``analytic priors,'' 
even though 
there are no closed expressions 
for the EFT parameters
that one can use. 

Our results are presented in 
figs.~\ref{fig:dist_QH}~\ref{fig:dist_QH_ctr}
and in 
table~\ref{tab:stats1},
where we show the mean values
and standard deviations
of the EFT parameters.

We start our comparison 
with the bias parameters.
The density of our 
analytic HOD samples
and the samples from the
simulations 
for the same set of HOD 
parameters are displayed 
in fig.~\ref{fig:dist_QH}.
Overall, our analytic model works
for the bias parameters quite well, estimating 
the 1D marginal distributions within $\lesssim 40\%$.
We find the largest discrepancy at the level of $b_3$. 
However, since this parameter
does not appear in the one-loop
power spectrum model, it is irrelevant 
for actual data analyses
based on the tree-level
bispectrum e.g.~\cite{Philcox:2021kcw,Ivanov:2023qzb,Ivanov:2021kcd}. 

The stochasticity
parameters and 
RSD counterterms 
are presented 
in fig.~\ref{fig:dist_QH_ctr}.
We observe highly 
consistent results 
for the redshift space counterterms. 
In particular, the 1D mean and standard 
deviations of $b_{4}$ and $c_{\mu^2}$
are reproduced to within $10\%$. 

Importantly, our simple analytic model
for the deterministic EFT parameters (bias 
parameters and RSD counterterms) 
reproduces all 
non-Gaussian correlations between them. 

We find 
consistent results
for the stochasticity 
EFT parameters $\alpha_0,\alpha_1,\alpha_2$,
but the good agreement
in this case is to a large extent 
thanks for fudge factors 
that we have calibrated  
to reproduce the 
target standard deviations
of the HOD priors. 
Even with these additional fudge factors
the agreement is typically within $\lesssim 40\%$.
While this error is acceptable for the generation
of EFT priors, it clearly calls 
for the development 
of more realistic models for the stochastic 
part of the galaxy 
power spectrum.

\begin{table}[ht!]
\centering
\begin{tabular}{|c|c|c|}
\hline
Param. & Analytic & Simulation \\
\hline
$b_1$ & 2.19 $\pm$ 0.56 & 2.05 $\pm$ 0.49 \\
$b_2$  & 0.62 $\pm$ 1.02 & 0.37 $\pm$ 0.82 \\
$b_3$ & 0.14 $\pm$ 1.65 & -0.38 $\pm$ 1.21 \\
$b_{\G}$ & -0.56 $\pm$ 0.64 & -0.46 $\pm$ 0.47 \\
$b_{\GG}$  & 1.53 $\pm$ 2.42 & 1.14 $\pm$ 1.81 \\
$b_{\nabla^2\delta}$ & 2.62 $\pm$ 0.68 & 2.19 $\pm$ 0.67 \\
$\alpha_0$ & -0.12 $\pm$ 0.20 & -0.29 $\pm$ 0.20 \\
$\alpha_1$ & 0.19 $\pm$ 0.17 & 0.14 $\pm$ 0.19 \\
$\alpha_2$ & 0.26 $\pm$ 1.53 & 0.51 $\pm$ 1.36 \\
$c_{\mu^2}$ & -46.99 $\pm$ 34.93 & -40.93 $\pm$ 31.40 \\
$b_4$ & 290 $\pm$ 648 & 282 $\pm$ 686 \\
\hline
\end{tabular}
\caption{Mean values and standard deviation of the EFT parameter samples
from our analytic approach and simulations generated
for the same range of HOD parameters
for the \textit{Planck2018} cosmology~\cite{Planck}.
$b_{\nabla^2\delta}$ and $c_{\mu^2}$ 
are quoted in units $[\Mpch]^2$, while 
for $b_4$ we use units of $[\Mpch]^4$.
}
\label{tab:stats1}
\end{table}

Note that the agreement between the simulations
and the deterministic EFT parameters is 
a highly non-trivial test 
of our analytic approach, as well as the simulation-based
field-level priors derived in~\cite{Ivanov:2024hgq,Ivanov:2024xgb}. 
Indeed, if our field-level fits 
for the \textit{halo} bias dependencies $b_{\mathcal{O}_a}^h(\nu)$ 
had been biased by two-loop corrections 
(which is equivalent to a wrong choice of $\kmax$),  
we would have obtained biased analytic  
predictions for $b_{\mathcal{O}_a}^g$
that would not match the parameters measured
from the simulations.\footnote{Note that the two-loop corrections of halos and HOD galaxies are different, so if present, they would be expected to bias 
halo and HOD measurements 
differently.}

All in all, we conclude that 
our analytic approach based on the halo model
leads to an excellent agreement with 
the N-body data 
for galaxy bias parameters
and counterterms, and satisfactory 
results for the stochastic EFT parameters.

\section{Testing cosmology 
and HOD dependence
of the EFT priors}
\label{sec:cosmo}

Let us discuss now the cosmology 
and HOD 
dependence of the EFT priors. 

\subsection{Analytic study}

Let us first discuss the cosmology
dependence of the galaxy EFT parameters,
following from the analytic equation~\eqref{eq:biasHOD}. 
This will allow us to gain insights 
into the cosmology-dependence 
of EFT parameters at the qualitative level,
and to some degree 
at the quantitative level. 

Let us fist discuss galaxy bias. 
It is useful to switch the integration variable in the HOD integrals from the mass to the peak height parameter $\nu$, in which case the relevant integral 
takes the form
\be 
\label{eq:biasHOD_nu}
b^g_{\mathcal{O}_a}
=\frac{1}{\bar n_g}\int d\nu \frac{d \bar n(\nu)}{d\nu}\langle N_g\rangle_M
b^h_{\mathcal{O}_a}(\nu)\,,
\ee 
where $\frac{d \bar n(\nu)}{d\nu}$ is given by:
\be 
\frac{d \bar n(\nu)}{d\nu} =
\frac{d \bar n}{d \ln M}\frac{d\ln M}{d\nu}
=\frac{\bar \rho_m(t_0)}{M}f(\nu)~\,.
\ee 
In order to obtain closed analytic 
expressions, let us assume a
power-law cosmology.
The linear matter power spectrum 
is given by
\be 
P_{11}(k)=\frac{2\pi^2}{k^3_{\rm NL}} \frac{k^n}{k_{\rm NL}^{n}}\,,
\ee 
where $k_{\rm NL}$
is the non-linear scale. By definition, 
it is the scales at which the 
position space density variance
becomes of order unity, 
which is particularly easy to see from the above definition. Note that $k_{\rm NL}$
has a non-trivial redshift-dependence
which can be inferred from the 
linear theory scaling
$P_{11}\propto D^2_+(z)$.
The peak height in the power
law cosmology is given by 
\be
\label{eq:map}
\begin{split}
&\nu = \delta_c \left(\frac{M}{\tilde{M}_{\rm NL}}\right)^{\frac{n+3}{6}}\equiv \left(\frac{M}{M_{\rm NL}}\right)^{\frac{n+3}{6}}\,,\\
&M=M_{\rm NL}\nu^{\frac{6}{n+3}}\,,
\end{split}
\ee
where we used 
\be
\begin{split}
&\sigma_M^2(z)  
=\left({\tilde{M}_{\rm NL}}/{M}\right)^{\frac{n+3}{3}}\,,\\
&\tilde{M}_{\rm NL}\equiv 
\left(\frac{9(1+n)\Gamma(n-1)\sin\left(\frac{\pi n}{2}\right)}{2^{n}(n-3)}\right)^{\frac{3}{n+3}}\frac{4\pi \Omega_{m,0} \rho_c}{3 k^3_{\rm NL}(z)}\,,
\end{split}
\ee 
and $4\pi R^3\rho_c \Omega_{m,0}/3=M$. This implies 
that\footnote{Note that the critical density
is cosmology-independent in the appropriate units,
$\rho_c=2.77\cdot 10^{11}h^2M_\odot$Mpc$^{-3}$.}
\be 
\label{eq:biasHOD_nu2}
b^g_{\mathcal{O}_a}
=\frac{\int d\nu f(\nu) \nu^{-\frac{6}{n+3}}\langle N_g\rangle_{M[\nu]}
b^h_{\mathcal{O}_a}(\nu)}{\int d\nu f(\nu) \nu^{-\frac{6}{n+3}}\langle N_g\rangle_{M[\nu]}}\,.
\ee 
We observe that there are two sources
of cosmology-dependence: the factor 
$\nu^{-\frac{6}{n+3}}\propto M^{-1}$
from the HMF and the HOD
dependence of $M$ 
whose mapping to $\nu$ 
is cosmology-dependent. 
Using the standard HOD parametrization~\eqref{eq:HODs},
eq.~\eqref{eq:biasHOD_nu2}
can be rewritten as 
\be 
b^g_{\mathcal{O}_a}=\frac{1}{\mathcal{N}} \left(b^g_{\mathcal{O}_a}\Big|_{c}+b^g_{\mathcal{O}_a}\Big|_{s}\right)\,,
\ee 
where
\be 
\label{eq:bias_gal_via_nu_final}
\begin{split}
&b^g_{\mathcal{O}_a}\Big|_{s}=
\int d\nu f(\nu) \nu^{-\frac{6}{n+3}}\langle N_c \rangle \left(\frac{\nu^{\frac{6}{n+3}}-\kappa \nu^{\frac{6}{n+3}}_{\rm cut}}{\mathcal{M}_1}\right)^{\alpha}
b^h_{\mathcal{O}_a}(\nu)\,,\\
&b^g_{\mathcal{O}_a}\Big|_{c}=
 \int d\nu f(\nu) \nu^{-\frac{6}{n+3}}\langle N_c \rangle_\nu b^h_{\mathcal{O}_a}(\nu) \,, 
\end{split}
\ee 
and we introduced
\be 
\begin{split}
&\mathcal{N}=\int d\nu f(\nu) \nu^{-\frac{6}{n+3}}\langle N_g\rangle \,,\\
&\langle N_c \rangle_\nu =
\frac{1}{2}\left[1+\text{Erf}\left(\frac{\log\nu-\log\nu_{\rm cut}}{\sqrt{2}\sigma'}\right)\right]\,,
\\
&\nu_{\rm cut} = \frac{\delta_c}{\sigma_{M_{\rm cut}}}=
\left(\frac{M_{\rm cut}}{M_{\rm NL}}\right)^{\frac{3+n}{6}}\,,
\end{split}
\ee 
and $\mathcal{M}_1={M}_1/M_{\rm NL}$, $\sigma'=6\sigma/(n+3)$. We see that the bias parameters of galaxies depend on the cosmological parameter
$n$,
the parameters $\nu_{\rm cut}$, $\sigma'$,
$M'_1$ which mix cosmological and HOD parameters,
and the HOD parameters $\alpha$ and $\kappa$.

Let us first study the 
dependence of the HOD on the amplitude
of the matter power spectrum at a redshift $z$,
parameterized by $\sigma_8(z)$. 
Consider two cosmologies with different 
values of this parameter, $\sigma_8(z)$
and $\sigma'_8(z)$. Changing cosmology
from $\sigma_8(z)$ to $\sigma'_8(z)$
is is equivalent to the re-scaling of 
$\nu_{\rm cut}$ and $M'_{1}$ as
\be 
\label{eq:nucutabs}
\begin{split}
&\nu_{\rm cut}\to \nu'_{\rm cut} = \nu_{\rm cut}\frac{\sigma_8(z)}{\sigma'_8(z)}\,,\\
& \mathcal{M}_1 \to \mathcal{M}'_1=\mathcal{M}_1\left(\frac{\sigma_8}{\sigma'_8}\right)^{\frac{6}{n+3}}
\end{split}
\ee 
Note that 
$\sigma'$, $\alpha$, $\kappa$ do not change 
in this case. The shift of \eqref{eq:nucutabs}
due to $\sigma_8$ is equivalent to 
shifts in $\nu_{\rm cut}$ 
and $\mathcal{M}_1$ due to the change of the threshold halo masses
$M_{\rm cut}$ and $M_1$. In other words, 
the cosmology dependence is degenerate 
with the values of the HOD 
parameters themselves. To see that, let us require 
the same HOD as a function of $\nu$ 
in both cosmologies. This implies that 
\be 
\begin{split}
& \nu_{\rm cut}=\nu'_{\rm cut}\Rightarrow 
M'_{\rm cut}=M_{\rm cut}\left(\frac{\sigma'_8}{\sigma_8}\right)^{\frac{6}{n+3}}\,,\\
& \mathcal{M}'_1 =\mathcal{M}_1 \Rightarrow  M_1=M_1 \left(\frac{\sigma'_8}{\sigma_8}\right)^{\frac{6}{n+3}}~\,.
\end{split}
\ee 
Using $n=-2$, $\sigma'_8 = 0.7$,
and $\sigma_8=0.8$ we find that 
\be 
\log M'_{\rm cut}-\log M_{\rm cut}=
\log M'_{1}-\log M_{1}\approx -0.3\,.
\ee 
Hence, a $10\%$ shift in the amplitude 
of the linear matter power spectrum
at a given redshift can be 
fully compensated by a relatively small
shift of the threshold halo masses. The sign of the difference also makes sense: in a cosmology 
with weaker matter fluctuations, one has 
to decrease the threshold halo mass
in order to maintain the same abundance 
of galaxies. 

Note that our argument about the degeneracy 
between the linear mass fluctuation
amplitude and the HOD parameters is, 
in fact, exact: one can absorb the $\sigma_8(z)$
dependence of eqs.~\eqref{eq:bias_gal_via_nu_final}
into the HOD parameters completely. The transformations
required for that are simply
\be 
\begin{split}
\nu_{\rm cut} \to \nu_{\rm cut}\frac{\sigma'_8(z)}{\sigma_8(z)}\,,\quad \mathcal{M}_1\to \mathcal{M}_1 \left(\frac{\sigma'_8}{\sigma_8}\right)^{\frac{6}{n+3}}\,.
\end{split}
\ee 
In the general $\Lambda$CDM case the rightmost 
transformation above will require an inversion of the 
function $\sigma_M(z)$, which 
can be easily done numerically. 

The situation becomes more complicated
once we consider the dependence on the effective tilt
of the matter power spectrum $n$. 
As we can see from eqs.~\eqref{eq:bias_gal_via_nu_final},
it is impossible to eliminate this dependence 
completely even in the simplified 
example of the power-law 
cosmology. Therefore, we expect some 
degree of residual cosmology dependence.
Quantifying this is one of the goals of our work. 

As far as the RSD counterterms are concerned, 
the part inherited from the halos 
exhibits the same dependence
on cosmology as the bias parameters, 
and therefore our conclusion
about the approximate cosmology
dependence is true for them. The additional 
terms that depend on the velocity bias
are cosmology-dependent as well, 
but their amplitude is fully 
degenerate with the 
velocity bias parameters, and therefore, 
this cosmology dependence can be absorbed exactly. 
Note, however, that the higher order bias 
$b_{\nabla^2\delta}$
and the RSD counterterm $c_{\mu^2}$
also depend on the dark matter counterterms
which are cosmology-dependent. For instance,
\be 
b_{\nabla^2\delta}=b'_{\nabla^2\delta}-b_1c_s\,,
\ee 
where $c_s$ is the cosmology-dependent 
speed of sound of dark matter. 
In principle, the cosmology dependence
of $c_s$ can be estimated using semi-analytic
arguments along the lines of~\cite{Chudaykin:2022sdl,Nascimento:2024rbv,Baldauf:2015tla,Baldauf:2015zga}.
However, for the purpose of our work, 
it will be sufficient to use the naturalness arguments,
according to which $c_s$ should be of the order
of the error that we make by evaluating all
EFT integrals up to the infinite momenta. 
It implies the estimate 
\be 
c_s\sim \frac{61D^2_+(z)}{1260\pi^2}\int_{ k_{\rm NL}}^\infty dq~P_{11}(q)~\,,
\ee 
where $k_{\rm NL}\sim 0.5~\hMpc$ at $z=0.5$. 
A variation of the linear matter power spectrum
by $\sim 10\%$ would only lead to a similar 
$\sim 10\%$  variation of $c_s$. This parameter by itself 
is a small part of $b_{\nabla^2\delta}$ ($c_s(z=0.5)\simeq 0.5~[\Mpch]^2$), which changes by a factor of few under the variation of
HOD parameters. Thus, we conclude that 
the cosmology-dependence of 
the dark matter counterterms can be safely ignored. 

Finally, let us comment on the cosmology dependence
of the stochastic EFT parameters. Using the HOD expressions
 \eqref{eq:a0_HOD}
 and
\eqref{eq:a1_naive},
and repeating our 
arguments for the galaxy bias, 
we can conclude that 
the dependence on 
$\sigma_M(z)$
can be absorbed 
into the HOD parameters exactly, but the 
dependence on the matter
power spectrum slope will remain. 

Let us now present
explicit numerical
results that will quantify
the residual cosmology 
dependence within our 
analytic model. 

\subsection{Numerical results}

To obtain numerical results
for different cosmologies,
we prepare a latin 
hypercube of cosmological parameters $\omega_b,\omega_{cdm}$, $n_s$, $h$,
and $\sigma_8$. We sample 
these parameters randomly from the 
following flat distributions
\be 
\begin{split}
& \omega_{cdm} \in [0.1 , 0.132]\,,\quad 
\omega_{b} \in [0.0220,0.0228]\,,\\
& h\in [0.65, 0.75]\,,
 n_s \in [0.9,1.01]\,,\quad 
 \sigma_8\in [0.6, 0.9]\,.
\end{split}
\ee 
We keep the total 
neutrino mass
to be fixed to its minimally allowed value 
$0.06$~eV. 
For parameters $\omega_{cdm}$ and $n_s$
our priors correspond to 
approximately $10\times$ the 
width of 
the \textit{Planck2018}
standard deviations~\cite{Planck}. 
For 
$\omega_b$ we use only one standard deviation as this 
parameter is determined
from the CMB in a nearly
model-independent way. 
The wide ranges of $h$
and $\sigma_8$ we use are motivated by the 
so-called $H_0$
and $\sigma_8$ tensions~\cite{Abdalla:2022yfr}.

\begin{figure*}[ht!]
\centering
\includegraphics[width=1.00\textwidth]{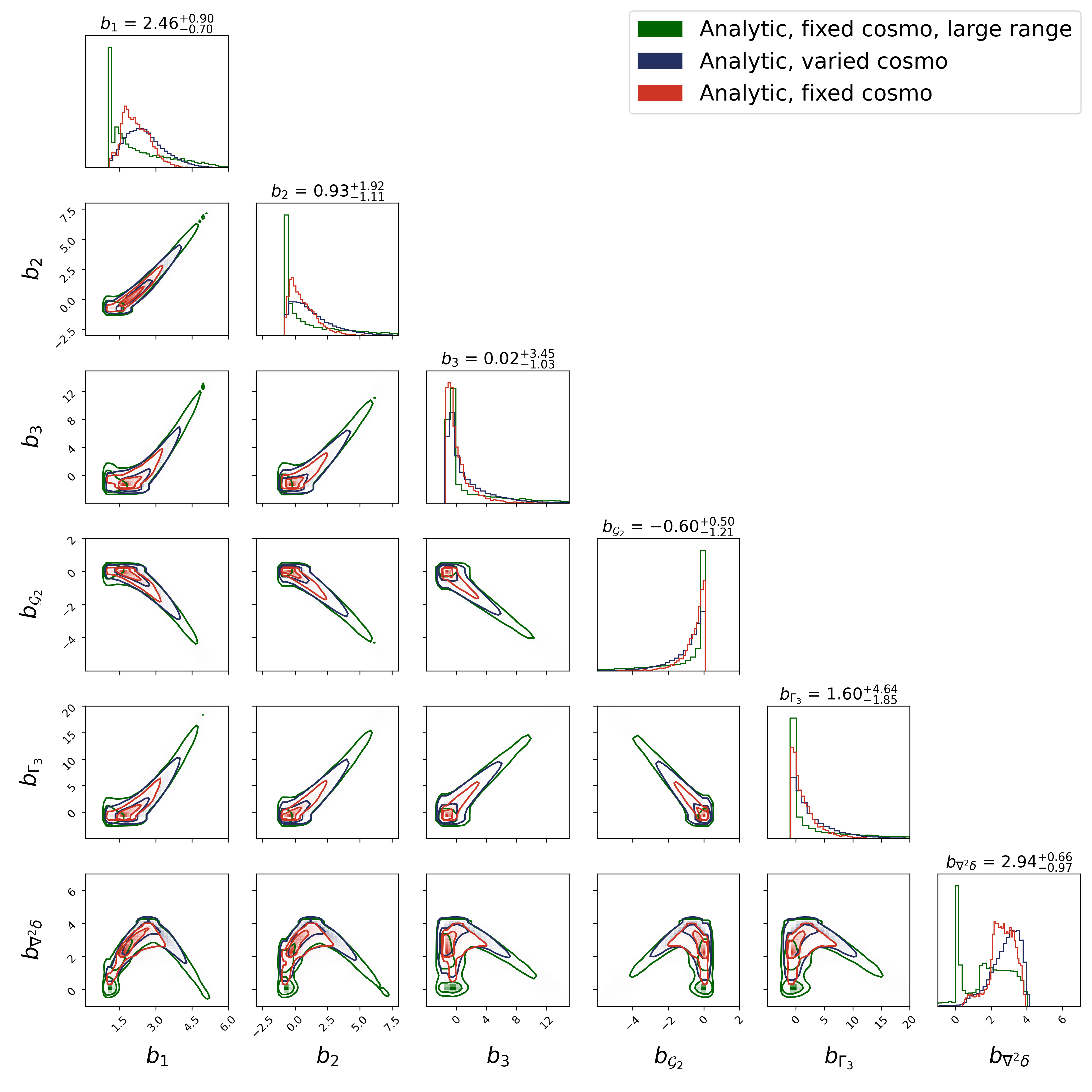}
   \caption{The distribution of bias EFT parameters
   from our analytic model. We display the sample
   produced for a fixed cosmological model,
   and a wide range of the HOD parameters (green), 
   the sample produced for the usual range of HOD parameters and by varying cosmology (blue),
   and the baseline sample produced
   at a fixed cosmology (red).
Density levels correspond to two-dimensional $1$-$\sigma$
  and $2$-$\sigma$ intervals (i.e. 39.3\% and 86.5\% of samples). 
    } \label{fig:dist_all_bias}
\end{figure*}

\begin{figure*}[ht!]
\centering
\includegraphics[width=1.00\textwidth]{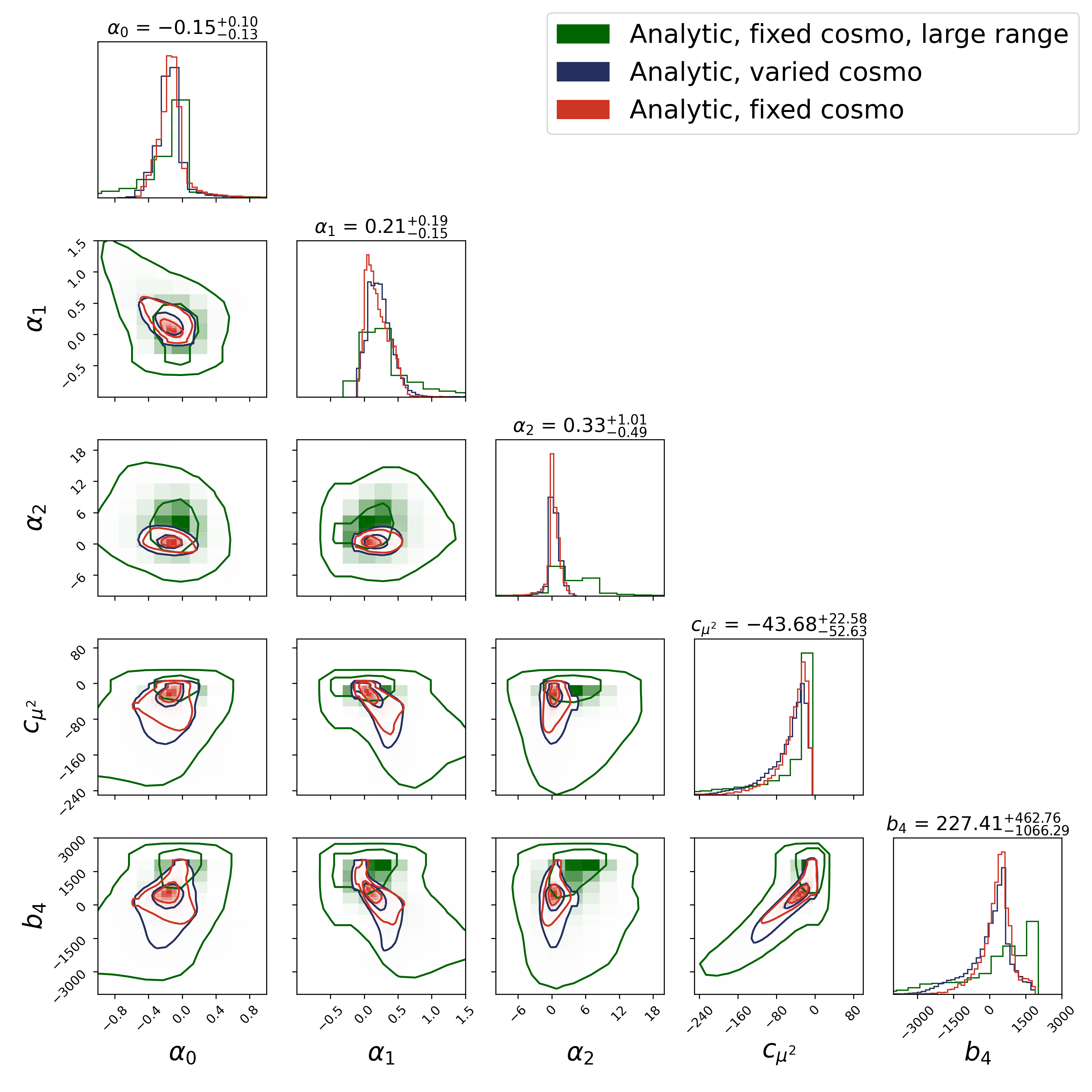}
   \caption{Same as fig.~\ref{fig:dist_all_bias}
   but for the stochastic 
   EFT parameters and redshift-space
   counterterms. 
    } \label{fig:dist_all_ctr}
\end{figure*}

\begin{table*}[htb!]
\centering
\begin{tabular}{|c|c|c|c|c|c|}
\hline
Param. & Fixed cosmo  & Varied cosmo & Large range, fixed cosmo & 
Large range, Sim. 
& Decorated HOD (Sim.) \\
\hline
$b_1$ & 2.1898 $\pm$ 0.5624 & 2.5538 $\pm$ 0.7972 & 2.2778 $\pm$ 1.2808 
& 2.1384 $\pm$ 1.1591
& 2.1260 $\pm$ 1.4268 \\
$b_2$ & 0.6228 $\pm$ 1.0281 & 1.3060 $\pm$ 1.5831 
& 1.1001 $\pm$ 2.5334 
& 0.8188 $\pm$ 2.1090
& 0.8584 $\pm$ 2.1826 \\
$b_3$ & 0.1388 $\pm$ 1.6579 & 1.1036 $\pm$ 2.8270 & 1.6913 $\pm$ 4.7106 
& 1.0898 $\pm$ 4.0851
& 1.0881 $\pm$ 3.7282 \\
$b_{\G}$ & -0.5652 $\pm$ 0.6373 & -0.9299 $\pm$ 1.0144 & -0.9859 $\pm$ 1.7383 
& -0.8240 $\pm$ 1.4524
& -0.9219 $\pm$ 2.0868 \\
$b_{\GG}$ & 1.5299 $\pm$ 2.4241 & 2.9072 $\pm$ 3.8740 
& 3.2070 $\pm$ 6.6383 
& 2.6909 $\pm$ 5.7106
& 3.0875 $\pm$ 8.4037 \\
$b_{\nabla^2\delta}$ & 2.6254 $\pm$ 0.6833 & 2.7869 $\pm$ 0.8382 
& 1.6988 $\pm$ 1.3697 
& 1.8742 $\pm$ 1.1757
& 1.2832 $\pm$ 6.3323 \\
$\alpha_0$ & -0.1166 $\pm$ 0.2022 & -0.1469 $\pm$ 0.1884 
& -0.1481 $\pm$ 0.4954 
& -0.0339 $\pm$ 0.8873
& 0.2653 $\pm$ 1.3830 \\
$\alpha_1$ & 0.1996 $\pm$ 0.1742 & 0.2293 $\pm$ 0.1795 
& 0.3914 $\pm$ 0.5760 
& 0.1275 $\pm$ 0.2743
& 0.0407 $\pm$ 0.2791 \\
$\alpha_2$ & 0.2585 $\pm$ 1.5363 & 0.3496 $\pm$ 1.4508 
& 3.5524 $\pm$ 6.4653 
& -0.1371 $\pm$ 6.6088
& -1.5075 $\pm$ 10.7245 \\
$c_{\mu^2}$ & -46.9907 $\pm$ 34.9312 & -57.4220 $\pm$ 44.0448 
& -72.2467 $\pm$ 111.7737 
& -65.4102 $\pm$ 101.8349
& -65.1294 $\pm$ 100.2469 \\
$b_4$ & 290.1989 $\pm$ 648.1431 & 8.8693 $\pm$ 886.7947 
& -62.0217 $\pm$ 2413.4284 
& -10.8034 $\pm$ 2000.4312
& 58.3832 $\pm$ 2029.0751 \\
\hline
\end{tabular}
\caption{Mean and standard deviation of EFT parameter
samples generated with our analytic approach 
assuming a fixed cosmology (first column), 
varied cosmology (second column),
or a fixed cosmology and an extended range
of the HOD parameters (third column). 
For comparison, we also show the simulation
results for the same large range of the HOD parameters derived within the base HOD model (fourth column),
and for the decorated HOD models that incorporate
effects of assembly bias
and baryonic feedback (fifth column). 
}
\label{tab:stats}
\end{table*}

\begin{figure*}[ht!]
\centering
\includegraphics[width=1.00\textwidth]{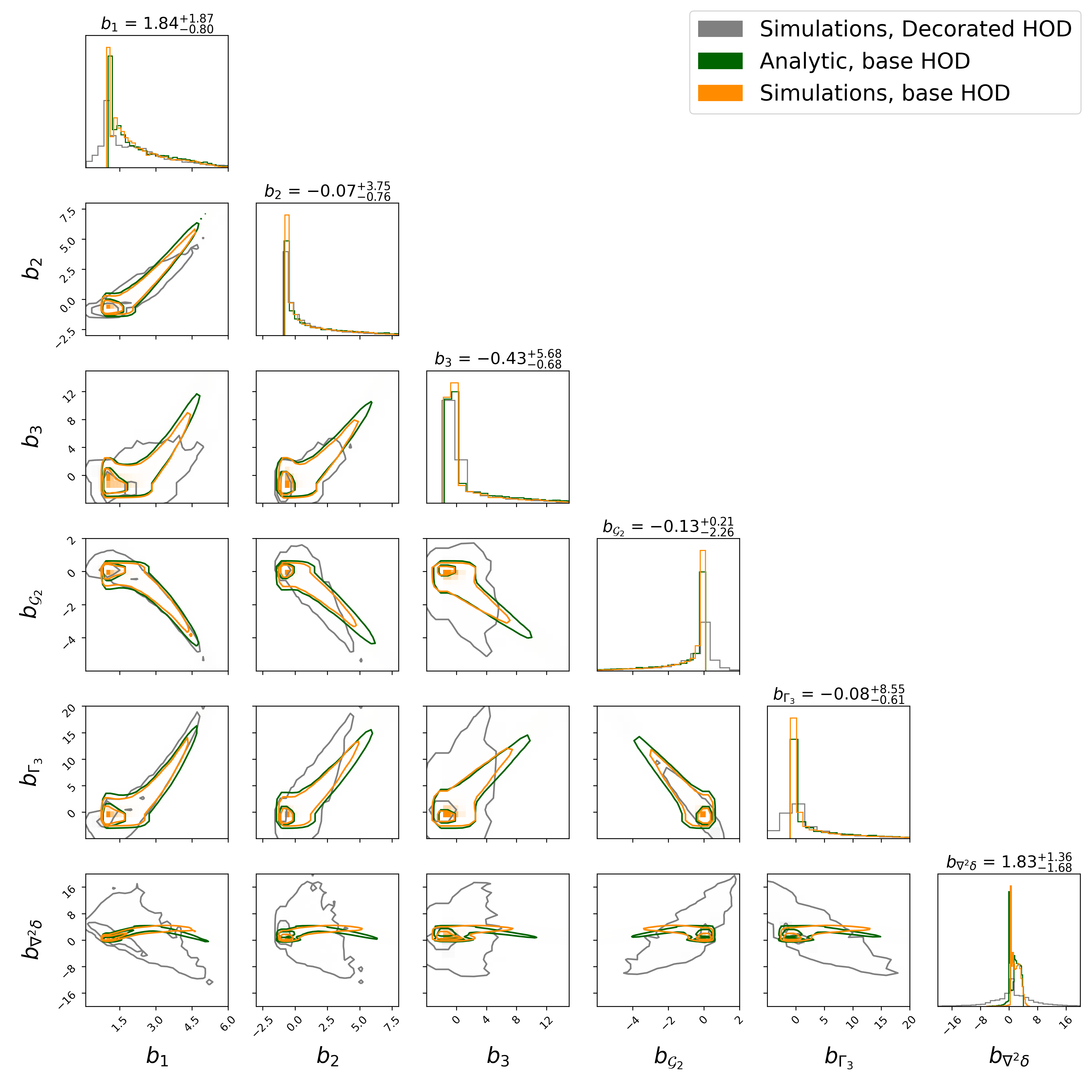}
   \caption{The distribution of bias EFT parameters
   from our analytic model (dark green),
   the simulated base HOD catalogs (orange),
   and the simulated decorated HOD catalogs 
   (gray).
    } \label{fig:comp_bias}
\end{figure*}

\begin{figure*}[ht!]
\centering
\includegraphics[width=1.00\textwidth]{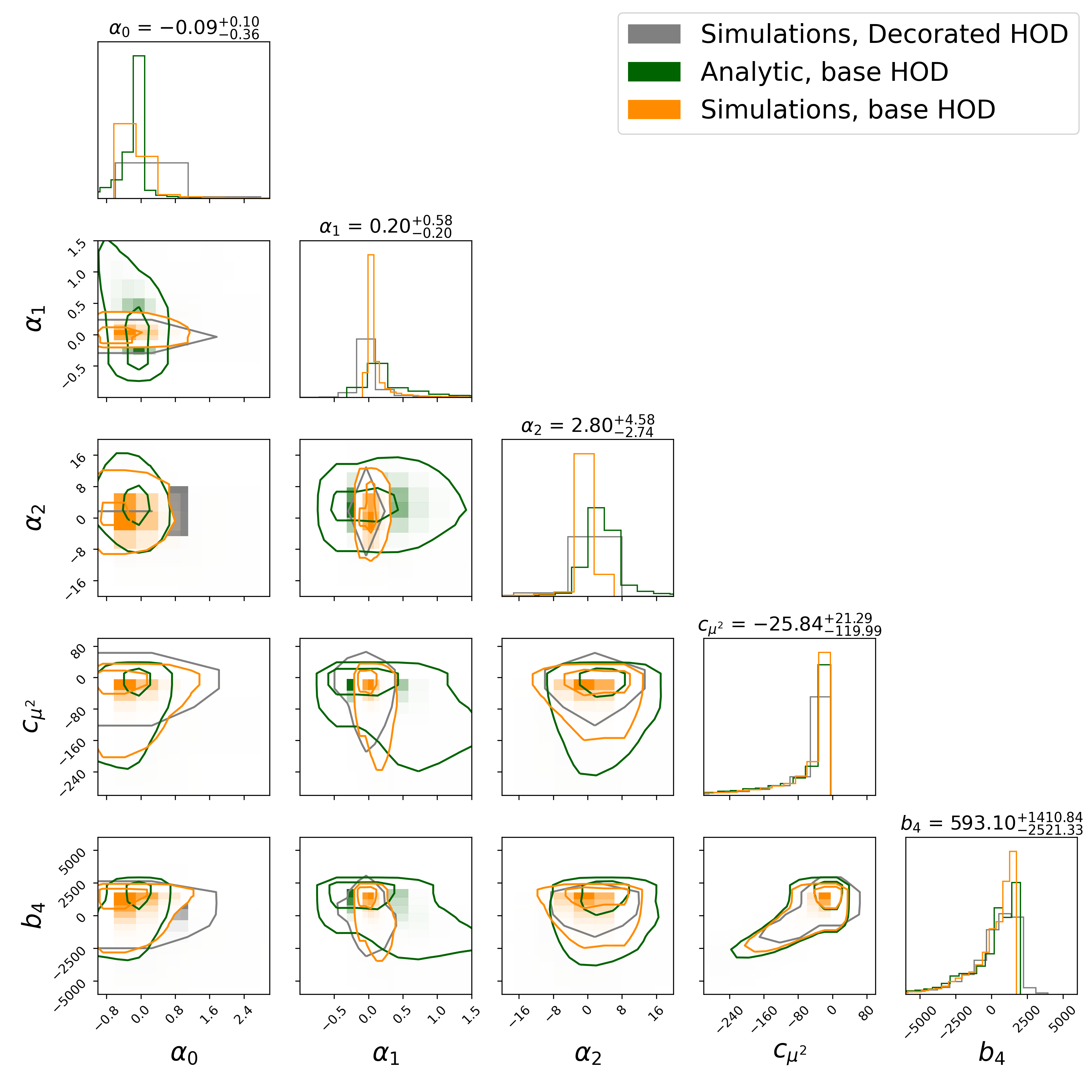}
   \caption{Same as fig.~\ref{fig:comp_bias}
   but for the stochastic 
   EFT parameters and redshift-space
   counterterms. 
    } \label{fig:comp_rsd}
\end{figure*}

In the cosmology-dependent sample
we use the same priors on the HOD
parameters as we used before~\eqref{eq:lrghod1}. To test
the assertion that the cosmology-dependence is degenerate with the HOD parameters, we compare our cosmology
dependent EFT parameters 
samples with those samples from
HOD models at a fixed cosmology, 
but with wider priors on HOD parameters. To that end we use
the following HOD priors 
motivated by~\cite{Ivanov:2024xgb,DESI:2023ujh}:
\be 
\label{eq:extHOD_pr}
\begin{split}
& 
\log_{10} M_{\rm cut}\in [12,14]\,,\quad 
\log_{10} M_1\in [13,15]\,,\\
& \log \sigma \in [-3.5,1.0]\,,\quad  \alpha \in [0.5,1.5]\,,\quad \\
& \alpha_c \in [0,1]\,,
\quad 
\alpha_s \in [0,2]\,,
\quad  \kappa \in [0.0,1.5]\,.
\end{split}
\ee 

Our results are presented in figs.~\ref{fig:dist_all_bias},~\ref{fig:dist_all_ctr} for the bias
parameters and counterterms,
respectively. For comparison, 
we also display the analytic samples
at a fixed cosmology
with the baseline HOD priors
in this plot.
The 1D intervals of the EFT parameter
samples are presented in table~\ref{tab:stats}. 
For comparison, we also quote the simulation
results for the same HOD priors~\eqref{eq:extHOD_pr}, 
as well as the simulation-based results
from~\cite{Ivanov:2024xgb}
for HOD models that include the decorated 
modifications. The corresponding 
distributions are displayed in 
figs.~\ref{fig:comp_bias},~\ref{fig:comp_rsd}.

Our results show that
the variation of the HOD parameters
produces a spread of the EFT parameters that is much wider than
the one generated by the variation
of cosmology. Comparing the fixed cosmology samples
with the varied cosmology samples obtained 
under the same HOD priors,
we note that the cosmology 
dependence only weakly affects the tails of the distribution,
and therefore, at a leading order,
it can be ignored altogether. 

As a consistency check, we have compared 
our analytic results for EFT parameters
of the base HOD models with wide priors with 
the simulations for the same HOD parameters. 
Overall, we find a good agreement, except 
for the $\alpha_{0,1}$ parameters,
for which the width of the distribution
appears to disagree with the simulations
at about $100\%$. 
It is important to stress that our
fudge factors for the stochastic
parameters were calibrated 
to reproduce the standard deviations
of the HOD-based samples from the previous 
section, which were drawn from more
narrow priors. The discrepancy 
at the level of $\alpha_{0,1}$
is another call for a more realistic model. 
We note, however, that the width of the $\alpha_2$
parameter is reproduced quite well,
which implies that our analytic model
correctly captures the parametric 
dependence of $\alpha_2$ on the base HOD 
parameters.

As a final remark, let us notice
while our analytic model reproduces well
the base HOD results, the EFT parameters from the base HOD models in general
appear to be somewhat different 
than the EFT samples of the
decorated HOD models, 
see figs.~\ref{fig:comp_bias},~\ref{fig:comp_rsd}.
In particular, 
certain correlation between the 
bias parameters e.g. $b_2-b_1$ 
have a different slope. 
In addition, the decorated HOD model
in general produce a wider 
spread of the EFT parameters. This is 
especially so for the counterterm 
$b_{\nabla^2\delta}$. 
For the stochastic and redsfhit space
counterterms, however, the decorated HOD 
samples are quite similar to the usual base 
HOD results. 
All in all, figs.~\ref{fig:comp_bias},~\ref{fig:comp_rsd} display the obvious limitations
of the base HOD framework,
which is inherited in our analytic approach. 
This suggests that the use
of the base HOD to generate the priors
may lead to unrealistically narrow EFT 
priors, which will lead to biased
estimates of cosmological parameters. 

\section{Conclusions}
\label{sec:disc}

We have presented an analytic 
method to generate 
approximate simulation-based 
priors for the EFT-based 
full-shape analysis based
on the halo model framework. 
Our method builds on three key ingredients: (a) 
a universal analytic model for the halo mass function, 
(b) the standard 7-parameter HOD parametrization, and 
(c) the assumption that the halo EFT parameters
depend only on the peak height. 
In addition, we introduce a number
of fudge parameters that account for the failure of the standard analytic 
halo model to account for the 
measured values of the 
stochastic
EFT parameters. 

While our analytic method is based on the above assumptions, which are
quite restrictive, it offers many 
important practical advantages.
First, it gives an analytic 
insight into the dependence 
of the EFT parameters on the HOD
models and cosmology. 
In particular, we have proved 
analytically that the dependence 
of the EFT parameters on the amplitude
of the linear matter power spectrum
at a redshift of interest
is completely degenerate 
with the threshold halo 
mass values. In a more general 
setting, our method allowed us to generate a sample of EFT parameters
for a large grid of cosmological
models and demonstrate explicitly 
that the  
cosmology-independence of the 
EFT parameters within the HOD 
models
is negligible for the purpose
of the prior generation. 

A second important advantage of our 
approach is that it allows one to
easily 
estimate the EFT parameters of 
galaxies at different redshifts. 
For instance, it can be used 
to get insight into the EFT parameters
of quasars~\cite{Chudaykin:2022nru}
or the high-redshift galaxies
relevant for Spec-5
surveys~\cite{Ravi:2024twy}.

An important application
of our approach 
it the estimation of EFT
parameters in extended cosmological 
models. For instance, our approach
applies directly to models beyond $\Lambda$CDM
that modify the 
shape of the linear matter power 
spectrum before recombination~\cite{He:2023dbn,Camarena:2023cku,He:2023oke,He:2025npy,Ivanov:2020ril}, 
or scenarios that only change the background 
expansion rate, such as 
a non-zero spatial curvature or 
dynamical dark energies~\cite{Chudaykin:2020ghx}.
In principle, 
our method can also be used to
estimate EFT parameters in other 
extended cosmological models that 
modify the clustering 
at low redshifts, such as the 
ultralight axion dark matter~\cite{Rogers:2023ezo}
or massive light relics~\cite{Xu:2021rwg}.
We caution, however, that 
some of the assumptions
behind our approach, such as the 
universality of the halo mass mass 
function, may be violated
in such models. 

From the theory side, 
it will be important to 
develop a more realistic
model for the stochastic 
EFT counterterms 
that incorporates the exclusion effects along the lines of~\cite{Baldauf:2013hka,Sullivan:2021sof}.

Finally, it will be interesting 
to apply our approach to other
galaxy samples with 
different HOD parameterizations,
e.g. emission line galaxies~\cite{Alam:2019pwr,Ivanov:2024dgv}
or quasars~\cite{eBOSS:2020pip,Chudaykin:2022nru}.
We leave all the above research direction
for future work.

\section*{Acknowledgments}
We are grateful to Jamie Sullivan 
for useful conversations.

\bibliography{short.bib}

\end{document}